\documentclass[sigconf,authorversion]{acmart}
\renewcommand\footnotetextcopyrightpermission[1]{} % removes footnote with conference information in first column
\AtBeginDocument{%
  \providecommand\BibTeX{{%
    \normalfont B\kern-0.5em{\scshape i\kern-0.25em b}\kern-0.8em\TeX}}}
    
\usepackage{amsmath,amssymb,amsfonts} % check 
\usepackage{stmaryrd, bm} % check 
\usepackage{algorithmic} % check 
\usepackage{graphicx} % check 
\usepackage{textcomp} % check 
\usepackage{xcolor} % !!! not compatible, delete it later
\usepackage[ruled,vlined]{algorithm2e} % check
\usepackage{booktabs} % check 
\usepackage{comment} % !!! not compatible 
\usepackage{subcaption} %  !!! not compatible 
\usepackage{pifont} %  !!! not compatible 
\usepackage{multirow}  % check 
\usepackage{filecontents} %  !!! not compatible 
\usepackage{xspace} %  !!! not compatible 
\usepackage{wrapfig} %  !!! not compatible 
\usepackage{enumitem} %  !!! not compatible 

\setcopyright{acmcopyright}
\copyrightyear{2021}
\acmYear{2021}
\acmDOI{xxxxxxxxxxxx}
\acmConference[]{Paper under review}{}{}
\acmPrice{15.00}
\acmISBN{xxxxxxxxxxxx}

\SetKwInput{keyGeneration}{Key generation}
\SetKwInput{encryption}{Encryption}
\SetKwInput{decryption}{Decryption}
\SetKwInput{homoPro}{Homomorphic property}

\newcommand{\revisionok}[1]{{#1}}

\newcommand{\sysName}{KHOVID\xspace}
\newcommand{\redzone}{RedZone Collector\xspace}
\newcommand{\riskInter}{Risk Interpreter\xspace}

\begin{document}
%-------------------------------------------------------------------------------

\title{ KHOVID: Interoperable Privacy Preserving Digital Contact Tracing}

 \author{Xiang Cheng$^*$, Hanchao Yang$^*$, Archanaa S Krishnan $^*$, Patrick Schaumont$^\dagger$, and Yaling Yang$^*$ }

\affiliation{%
  \institution{Virginia Tech$^*$ \hspace{0.5em} Worcester Polytechnic Institute$^\dagger$}
}
\affiliation{%
  \institution{\{xiangcheng,hcyang,archanaa,yyang8\}@vt.edu$^*$,\hspace{0.3em}pschaumont@wpi.edu$^\dagger$  }
}

% \author{Xiang Cheng, Hanchao Yang, Archanaa S Krishnan }
% \affiliation{%
%   \institution{Virginia Tech}
% \email{xiangcheng@vt.edu,hcyang@vt.edu,archanaa@vt.edu}

% \author{Patrick Schaumont}
% \affiliation{%
%   \institution{Worcester Polytechnic Institute}
% \email{pschaumont@wpi.edu}

% \author{Yaling Yang}
% \affiliation{%
%   \institution{Virginia Tech}
% \email{yyang8@vt.edu}

% %for single author (just remove % characters)
% \author{
% {Xiang Cheng}\\
% Virginia Tech
% \and
% {Hanchao Yang}\\
% Virginia Tech\\

% \and
% {Archanaa S Krishnan}\\
% Virginia Tech\\

% \and
% {Archanaa S Krishnan}\\
% Virginia Tech\\

% \and{Patrick Schaumont}\\
% Worcester Polytechnic Institute\\

% \and
% {Yaling Yang}\\
% Virginia Tech
% } % end author

\begin{abstract}

During a pandemic, contact tracing is an essential tool to drive down the infection rate within a population. To accelerate the laborious manual contact tracing process, digital contact tracing (DCT) tools can track contact events transparently and privately by using the sensing and signaling capabilities of the ubiquitous cell phone. However, an effective DCT must 
not only preserve user privacy but also augment the existing manual contact tracing process. Indeed, not every member of a population may own a cell phone or have a DCT app installed and enabled. We present 
KHOVID to fulfill the combined goal of manual contact-tracing interoperability and DCT user privacy. At KHOVID's core is a privacy-friendly mechanism to encode user trajectories using geolocation data. Manual contact tracing data can be integrated through the same geolocation format. The accuracy of the geolocation data from DCT is improved using Bluetooth proximity detection, and we propose a novel method to encode Bluetooth ephemeral IDs. 
This contribution describes the detailed design of KHOVID; presents a prototype implementation including an app and server software; and presents a validation based on simulation and field experiments. We also compare the strengths of KHOVID with other, earlier proposals of DCT.
% This contribution describes the detailed design of KHOVID; presents a prototype implementation including an app and server software; and presents a validation based on a user survey, simulation, and field experiments. 

%\note{ Deadline: Deadline: 12 Dec 2020, 6:59:59 am EST Papers must be at most 12 pages including bibliography, a maximum of 3 additional pages can be used for well-marked appendices.}

\end{abstract}

%% >>>>>>>>>>> Change This <<<<<<<<<<<
% <concept_id>10002978.10003029.10011150</concept_id>
% <concept_desc>Security and privacy~Privacy protections</concept_desc>

% \ccsdesc[500]{Security and privacy~Privacy protections}
%% >>>>>>>>> Change this <<<<<<<<<<


\maketitle

\section{Introduction}
%% or give numbers
% \todo{Abundant recent evidences} in Singapore, South Korean and China have shown that a crucial tool in the fight with COVID-19 is smart-phone-based contact tracing that enables healthcare system to quickly identify individuals who had a close contact with confirmed cases \cite{strickland,Translating2020,UK2020,Cellphone2020}.  Such a mechanism can alarm these individuals so that they can voluntarily self-quarantine to prevent further spreading the disease.  In addition, when these individuals start to experience flu-like symptoms, contact tracing will identify them as likely COVID-19 cases and prioritize them for testing. This will help medical community to more effectively utilize the limited test capacity and provide early treatment for patients.  However, it has been a challenge to deploy such a system in US and many western countries due to strict privacy laws \cite{strickland,Translating2020,UK2020,Cellphone2020}, which prohibit the past whereabouts of both patients and general public to be collected and shared without affirmative consent.  

% Why is contact tracing important 

The global pandemic of COVID-19 has been a big challenge to human society.  By sacrificing economic growth, governments around the world are making efforts to control and mitigate its spreading. To effectively slow down the spreading of an infectious disease, the discovery rate of new infections should be considerably higher than the infection rate~\cite{eames2003contact}.  This requires contact tracing which identifies who was in close contact with diagnosed patients, so that  precautions can be taken to prevent further spread of the infection.
%Contact tracing was a vital tool for controlling the spread of SARS, MERS, and Ebola~\cite{ebolaContactTracing,SarsContactTracing} without shutting down social and economic activities.
 
% Contact tracing has been proven a vital tool for controlling the spread of viruses like SARS, MERS, and Ebola~\cite{ebolaContactTracing,SarsContactTracing} without shutting down social and economic activities.
% Its goal is to find who has close contact with diagnosed patients so that necessary precautions can be taken. Researchers~\cite{eames2003contact} also point out that to effectively slow down the spreading of an infectious disease,  the speed of discovering new infections should not be considerably lower than the infection rate.  This requires contact tracing to find possible contacts as fast as possible.  

Smartphone-based Digital Contact Tracing (DCT)  systems have been proposed in an attempt to  improve the speed of finding close contacts \cite{strickland,Translating2020,UK2020,Cellphone2020}. These systems
estimate pathogen exposure risk either  through geolocation (e.g. Global Navigation Satellite System (GNSS)), or via proximity-measurement technologies like Bluetooth proximity sensing. 
DCT systems can automatically notify these contacts so that they can self-quarantine to prevent further spreading of the disease.
In addition,  when these individuals start to experience sickness symptoms, DCT systems will identify them as likely infected cases and prioritize them for testing.  This will lead to more effective use of the limited test capacity and  early treatment for patients. %Recent research~\cite{abueg2020modeling,hinch2020effective} also shows that the adoption of DCT can meaningfully reduce infections, deaths, and hospitalizations. 

% the problem of current DCT systems 
% Nowadays, various 
Since the onset of COVID-19, various
DCT systems have been developed and deployed around the world. % digital contact tracing
 
%\revisionok{, including but not limited to }% such as TraceTogether~\cite{TraceTogetherWeb}, Corona-Warn~\cite{coronawarn} and Covid-Wise~\cite{googleAppleDCT,covidwise}. 
A majority of them % Among these deployed systems, most of them choose to 
identify contacts using Bluetooth rather than using GNSS due to privacy concerns. However, 
% we believe that 
if the privacy issue can be solved, geolocation-based DCT systems will offer 
two major advantages over Bluetooth-based systems.% important advantages over Bluetooth-based system in two critical aspects. 

First,
geolocation-based DCT systems are compatible with manual contact tracing, whereas Bluetooth-based systems are not.
% \revision{as they both deal with absolute location, whereas Bluetooth-based DCT systems use relative proximity information. } 
% which makes them more effective than Bluetooth-based systems. 
As of today, most of the contact tracing 
work is 
carried out manually. Contact tracers work with patients to identify their close contacts and 
the places they visited a few days before the onset of symptoms \cite{propublica}.
% the visited place. 
Apart from obtaining identities of close contacts from patients, manual contact tracing also identifies locations patients visited. This verifies contacts  whom the patient does not personally know, such as a cashier in a supermarket or a server in a restaurant~\cite{ContactTracingCDC,CoronavirusBBC}.
% Collecting visited locations can speed up contact tracing in several ways, like identifying contacts quickly in congregate  settings, or notifying special facilities (e.g., nursing home) to take quick actions~\cite{ContactTracingCDC,CoronavirusBBC}.
Unfortunately, Bluetooth-based  and other proximity-sensing-based DCT systems do not collect such absolute location data. Hence, they can neither benefit from nor aid manual contact tracing. 

%\note{Prof. Patrick: Emphasize the DCT is also important/helpful to manual contact tracing}
% the visited location information, which is absolute location data, from manual contact tracing cannot be used by Bluetooth-based systems as they use relative proximity based location information. }
% Unfortunately, both close contact and visited location information from manual contact tracing cannot be used by Bluetooth-based systems. 
In comparison, geolocation-based DCT systems and manual contact tracing can exchange location information for mutual benefits. For example, a geolocation-based DCT can provide exposure risk levels in public areas (e.g. restaurants and grocery stores) to help manual contact tracing to identify high-risk service workers. A contact tracer can share a patient's past visited locations to a geolocation-based DCT system even if the patient did not use the DCT app before the onset of symptoms. 
% In a geolocation-based DCT system, either contact tracing workers or patients can share their visited places to the DCT system, no matter whether they have downloaded the app before. 
By getting additional patient histories from manual contact tracing, geolocation-based DCT systems can warn more users 
and may achieve a reduction in disease spreading at a lower app adoption rate.
% thus
%require less adoption rate compared to Bluetooth-based systems to effectively 
%reduce % prevent
%the spreading of the disease. 
This makes geolocation-based DCT systems useful under a gradual deployment timeline, and in a demographic where a large portion of the population does not have the app. 
% in developing regions  or demographics where a large portion of the population do not own smartphones and can adopt a gradual deployment timeline. 
The latter scenario reflects reality. In the US for example, only 53\% of US seniors older than 65 own smartphones \cite{Timberg20most}. Yet, this age bracket is the most vulnerable to the disease.

Second, % Secondly, 
geolocation-based DCT systems can inform the health authorities about the spread of the disease in a particular location. By aggregation of patients’ trajectory history, the health authorities may learn the disease's spreading pattern within a community, and make informed decisions such as  
%
%The returns can help health authorities obtain the disease's spreading pattern within a community, and hence make 
%targeted % correct 
%decisions 
%such as
allocating more medical resources to hot-spot 
areas % places,
, announcing lockdown for a specific region instead of the whole state, or warn
% sending  warning 
about a recent high-risk activity to a local community. 
Unfortunately, % However, 
Bluetooth-based DCT systems cannot provide such information to the health authorities. 

\paragraph{Challenges:} Despite the above obvious benefits of geolocation-based DCT systems, they face two major design challenges:% Yet, designing a digital contact tracing system based on locations can be challenging. 
 % For example, 
% The recorded geolocations are sensitive information that can cause privacy concerns and the localization accuracy can be inaccurate in indoor environments. Here, 
%We summarize 
% two major challenges:
%  the major challenges into two points: 
\begin{enumerate}
    \item How can a DCT system  preserve all users’ privacy while recording their past trajectories? \item How to minimize the effects of GNSS localization error in contact tracing, especially in complex environments like indoor or urban canyon environments? 
\end{enumerate}

The first challenge  is critical since privacy concerns have greatly hindered the wide adoption  of DCT apps \cite{TimbergPrivacyConcern2020, LehmannPrivacyConcern2020}.  We consider 
% to preserve users' privacy from 
 two perspectives: 
(a) trajectory privacy that protects 
a user's visited places and any personal information inferred from the trajectory histories; 
(b) exposure privacy that hides the user's exposure risk provided by the DCT app. 
To protect users' trajectory privacy, we propose to use k-anonymity to mix users' real trajectories with synthesized trajectories, such
that users' real trajectories are hidden from  eavesdroppers and other parties in the DCT system~\cite{pappalardo2018data,bindschaedler2016synthesizing,sweeney2002k}. 
% For the second perspective, 
Next, we protect users' exposure risk through an efficient multi-party computation (MPC) design. MPC is a secure computation 
technique, where a function is computed by multiple parties jointly over their input, while keeping these inputs and the output private. However, designing an MPC-based DCT protocol is  nontrivial since MPC for general functions is currently not scalable due to its huge computational complexity~\cite{dou2016p}. To enable our MPC scheme to compute trajectory intersection efficiently even when a large number of 
synthesized trajectories exist, we customized the MPC scheme with a partial-homomorphic Paillier cryptosystem~\cite{paillier1999public}, where the computation of location intersection and exposure risk is decomposed
 % disintegrated 
 into a small set of homomorphic operations. 

For the second challenge, we propose to integrate the Bluetooth technology 
\revisionok{with our geolocation-based DCT system, }% into our DCT system, 
as Bluetooth has a better accuracy of proximity detection than GNSS in indoor environments. However, existing Bluetooth-based methods \cite{PEPP-PT, robert20,googleAppleDCT, troncoso2020decentralized} cannot fit into our geolocation-based design. 
%because they use unique random Ephemeral IDs (EphIDs)~, which may breach the k-anonymity \revisionok{of geolocation } and may leak the users' real trajectories to others. %\revisionok{For example, Raskar et al~\cite{raskar2020adding} suggested to add location context to GAEN API~\cite{googleAppleDCT}. Their intention 
%\revisionok{was }% is
%to make users aware of their exposure locations when getting exposure notifications, such that users can reject false-positives and self-assess their exposure. However, 
A naive integration of Bluetooth and geolocation-based DCT system may cause a privacy risk which may be exploited by curious users to deanonymize patients' identities using exposure locations. \revisionok{To prevent this problem, we designed our Bluetooth protocol to be compatible with our geolocation-based DCT system, which maintains k-anonymity based user privacy. 
}
% To solve this problem, we create a new Bluetooth protocol, which ensures trajectories’ k-anonymity will still be kept. 

% The reason is that intersections among real trajectories usually are accompanied by some exchange of EphIDs at the Bluetooth side, while fake trajectories' intersections do not have such records of EphIDs exchange.  For example, if Alice (patient) and Bob (uninfected) are two users of the geolocation-based DCT system, their recorded trajectory are protected under k-anonymity and uploaded to the central server for intersection check, yet the central server cannot distinguish their real trajectories from fake ones, however, if the EphIDs from existed work are uploaded together with the trajectories, the central server can check their intersections along with EphIDs, if two sets of trajectories intersect and EphIDs from Alice and Bob match at the same time, then both of two trajectories are real, otherwise at least one of the trajectory is fake. To solve this problem, we create a new Bluetooth protocol by introducing the location information to the EphIDs, which ensures trajectories’ k-anonymity will not be broken. 

Overall, in this paper, we 
propose \sysName: a \textit{\textbf{H}}ybrid DCT system that enhances geolocation-based contact tracing with Bluetooth while preserving user privacy using \textit{\textbf{K}}-anonymity.
% uses both geolocation- and Bluetooth-based location and preserves user privacy using \textit{\textbf{K}}-anonimity. 
% Overall, in this paper, we \revision{propose }% proposed a privacy-preserving geolocation-based DCT system, named \sysName.
In \sysName, degradation of user privacy 
% \revision{during } the integration 
is avoided by building a customized MPC mechanism using the Paillier cryptosystem.
In addition, to increase the accuracy of contact detection in complex environments, we combine the geolocation-based method with Bluetooth technology by designing a new Bluetooth DCT protocol. 
% \revision{It is a hybrid system that enhances the accuracy of contact detection by combining geolocaiton-based method with Bluetooth technology.}
\revisionok{\sysName offers accuracy and availability for multiple physical environments, interoperability with manual contact tracing, and the preservation of the privacy of all users in the system. }
% In this way, we ensure that \sysName can cooperate with manual contact tracing while ensuring accuracy and availability in different environments, and preserving all users’ privacy in the system simultaneously. 

\paragraph{Contribution:} We make the following contributions: 
\begin{itemize}[leftmargin=5pt]
    \item First, we proposed a geolocation-based contact tracing method that guarantees both trajectories and infection/exposure information are only known to users themselves. 
    
    \item Second, we designed a new Bluetooth-based DCT protocol that can coordinate with the geolocation-based method to boost contact tracing accuracy and availability of \sysName~. The novel design of EphIDs generation ensures that no additional personal information will be leaked compared to the geolocation-only method.
    
    \item Third, we implemented a prototype of \sysName, including both its app and server software. Field experiments were conducted to evaluate the effectiveness of \sysName system and showed that \sysName is highly scalable and can potentially slow and curb the spread of COVID-19 much faster than other DCT systems.
    
    % \item We also present preliminary results from a survey we conducted on the user perception of different design options in technology and privacy of DCT, demonstrating that \sysName's design will encourage more adoption of DCT.
\end{itemize}

\paragraph{Organization:}The rest of this paper is organized as follows.
Section \ref{sect:relatedwork} summarizes existing DCT designs.
Section~\ref{sec:sys-overview} provides an overview of \sysName's architecture and threat model. 
% Section~\ref{} describes our geolocation-based DCT system with its privacy protection techniques. 
Section~\ref{sec:BLE-method} describes our new Bluetooth-based DCT system and its hybrid operation with our geolocation-based DCT system described in Section~\ref{section:geo-method}. 
In Section~\ref{sec:sec_analysis}, we analyse the system security of \sysName , particularly against common attacks on DCT systems. 
Section~\ref{sec:evaluation} describes our simulation and field experiments on adoption rate and scalability of \sysName, respectively.
We conclude the paper by describing future work in Section~\ref{sec:conclusion}.

\section{Related Work}
\label{sect:relatedwork}
\begin{table*}[ht]

\caption{Comparison between the proposed \sysName system and related works. (PSI-Private Set Intersection, HE-Homomorphic Encryption, MPC - Multi-party computation)}

\resizebox{0.9\textwidth}{!}{%
\begin{tabular}{|ll|l|c|c|c|c|c|}
\hline
 &
  \textbf{Key Design Features}
   &
  \textbf{Protocol Name} &
  \textbf{\begin{tabular}[c]{@{}c@{}}Compatible \\ with manual \\ contact tracing\end{tabular}} &
  \textbf{\begin{tabular}[c]{@{}c@{}}Protect patients' \\ identity against \\ untrusting server\end{tabular}} & \textbf{\begin{tabular}[c]{@{}c@{}}Protect patients' \\ identity against \\ public \end{tabular}} &
  \textbf{\begin{tabular}[c]{@{}c@{}}Protect \\ exposure \\ risk\end{tabular}} &
  \textbf{\begin{tabular}[c]{@{}c@{}}Capture both direct\\  and indirect spreading \\  of infection\end{tabular}} \\ \hline
\multicolumn{1}{|l|}{\multirow{4}{*}{\rotatebox[origin=c]{90}{Bluetooth}}}         & \multirow{2}{*}{Centralized}   & TraceTogether~\cite{blueTraceWhitepaper}     &              &              &      $\bm{\checkmark}$     &   &  \\ \cline{3-8} 
\multicolumn{1}{|l|}{}                                   &                                & PEPP-PT~\cite{PEPP-PT}           &              &             & $\bm{\checkmark}$  &              &  \\ \cline{2-8} 
\multicolumn{1}{|l|}{}                                   & \multirow{2}{*}{Decentralized} & DP-3T~\cite{troncoso2020decentralized}             &              & $\bm{\checkmark}$ &              & $\bm{\checkmark}$   &          \\ \cline{3-8} 
\multicolumn{1}{|l|}{}                                   &                                & GEAN~\cite{googleAppleDCT}              &              &       $\bm{\checkmark}$ &        & $\bm{\checkmark}$ &              \\ \hline
\multicolumn{1}{|l|}{\multirow{5}{*}{\rotatebox[origin=c]{90}{Geolocation}}} & PSI                            & Berk et al.~\cite{berke2020assessing}       & $\bm{\checkmark}$ &         &     & $\bm{\checkmark}$ &               \\ \cline{2-8} 
\multicolumn{1}{|l|}{}                                   & \multirow{2}{*}{MPC}           & Fitzsimons et al.\cite{fitzsimons2020note} & $\bm{\checkmark}$ &              & & $\bm{\checkmark}$ &              \\ \cline{3-8} 
\multicolumn{1}{|l|}{}                                   &                                & Reichert et al.\cite{reichert2020privacy}   & $\bm{\checkmark}$ &              & & $\bm{\checkmark}$ &              \\ \cline{2-8} 
\multicolumn{1}{|l|}{}                                   & \begin{tabular}[c]{@{}l@{}}MPC+HE+\\ K-anonymity\end{tabular}                               & Proposed \sysName system        & $\bm{\checkmark}$ & $\bm{\checkmark}$ & $\bm{\checkmark}$ & $\bm{\checkmark}$ & $\bm{\checkmark}$ \\ \hline
\end{tabular}%

}

\label{tab:related_works}
\vspace{-0.1in}
\end{table*}
\revisionok{A majority of DCT systems for COVID-19 }
% Since the outbreak of COVID-19, many DCT methods have been proposed. The previous related works 
are  built on two technologies: Bluetooth and GNSS. The Bluetooth-based DCT exchanges Bluetooth beacons with nearby devices,  and the GNSS-based DCT traces users’ past visited places based on GNSS readings (i.e. latitude and longitude). In this section, we will briefly discuss related works based on these two types of contact tracing methods. The discussed limitations of these related works are summarized  in Table 1.

\subsection{Bluetooth-based Contact Tracing}

In a Bluetooth-based DCT, 
% after a user downloads the DCT app into his/her mobile device, 
an app on a user's mobile device generates unique randomized EphIDs. The app continuously broadcasts its EphIDs to neighboring devices, while saving
\revisionok{the received }% other devices’ advertised 
EphIDs from these neighbors. 
\revisionok{If }% Once
a user is tested positive with COVID-19, the sick user's app uploads the EphIDs that it has \revisionok{transmitted/received} in the past to a central server. Other users then can calculate their exposure risks by checking if their \revisionok{received/transmitted} EphIDs match the patients’ EphIDs. The exposure risks can be calculated either \revisionok{by }
% at 
the central server or by the 
users’ mobile phones, leading to a centralized or decentralized method, respectively (Table 1). 
%respectively
%(centralized method) or \revision{by the}
% at
%users’ mobile phones (decentralized method). 

\underline{\it Centralized methods} \revisionok{rely on central servers to calculate users' exposure risks and notify those with high exposure risk.}   TraceTogether~\cite{TraceTogetherWeb,blueTraceWhitepaper} in Singapore,  COVIDSafe in Australia~\cite{covidSafe}, and TousAntiCovid~\cite{PEPP-PT, tousanticovid, robert20} in France are examples of centralized design. However, the current centralized methods have raised many privacy concerns. For example, people are afraid that they may be forced into quarantine as the central server is aware of users’ exposure risks.  In addition, users’ social interactions  with their acquaintances are also exposed to the server 
% as it owns users’ encounter histories
\cite{reichert2020survey}. \revisionok{Such concerns of privacy have hindered the adoption of centralized schemes in many regions \cite{Timberg20most}}.

\underline{\it Decentralized methods} conduct exposure risk analysis on users’ mobile phones, by downloading patients’ EphIDs from the server.  DP-3T \cite{troncoso2020decentralized} and the Google/Apple Exposure Notification (GAEN) system~\cite {googleAppleDCT} are examples of decentralized methods. 
A few of 
\revisionok{the decentralized }% this type of 
apps \cite{covidwise,coronawarn,swisscovid,MICovidAlert}, deployed in small-scale, ensure that exposure risks are only known to users themselves. But 
\revisionok{they }% the methods 
also have limitations. These schemes reveal 
patients’ EphIDs and encounter time to all users in decentralized systems. This has been shown to easily leak the identities of patients~\cite{reichert2020survey,Vaudenay2020AnalysisOD,vaudenay2020centralized} and violate privacy protection for medical data.

%The choice between the centralized  and  decentralized methods is an unresolved problem. In contrast, our proposed method combines the advantages of both centralized and a decentralized methods such that users' privacy are protected against both other users and untrusted servers, regardless of whether the user is sick or healthy.

Besides security issues, Bluetooth-based DCT also  struggles with three additional performance challenges. The first one is poor performance under low to medium adoption rate \cite{bostonglobe,timemagazine}. %Besides compatibility challenges between the different schemes, a major challenge is to ensure that a sufficiently large number of people have an active contact tracing app. 
Bluetooth-based DCT only helps when both parties of a contact event have the app. If Alice with a Bluetooth-based DCT app sits in a public library besides a stranger who does not have the app, the encounter will not be captured by the system. As a result, the effectiveness of Bluetooth-based DCT under medium to low adoption rate is very poor. For example, if only 10\% of a population installs Bluetooth-based DCT in their phones, the chance that a random contact event can be captured by this app is only 1\%. 

%Besides the privacy concerns, there are also two important 
% effectiveness 
%drawbacks \revision{based on the effectiveness }of a standalone Bluetooth-based system. First, it only works when both persons in an encounter have the DCT app installed. Consider the case where person A with a Bluetooth-based DCT app sits besides a stranger B who does not have this app in a public library. This encounter will not be captured by Bluetooth-based DCT. Generally, the effectiveness of Bluetooth-based DCT under medium to low adoption rate is very poor. For example, if 10\% of a population installs Bluetooth-based DCT in their app, the chance that an encounter can be captured by this app is only 0.01.
\revisionok{A second major challenge }% In addition,
is that present Bluetooth-based DCT deployments are 
% Bluetooth-based DCT is 
incompatible with manual contact tracing. 
\revisionok{ Bluetooth EPhIDs transmitted by the sick user Alice are useless to a person without the app. The only way that the app-less person can be warned is by manual contact tracing.  }
% If Bob from the earlier example is later tested positive and manual contact tracing has identified all his past trajectories, Alice will not be informed about the high exposure risk.
 Bluetooth-based DCT can neither provide the location of dangerous contact events to manual contact tracing nor benefit from the location information obtained from manual contact tracing. Since manual contact tracing is the dominant tracing mechanism and is the only effective tracing mechanism for populations that do not own smartphones, this incompatibility limits the effectiveness of Bluetooth-based DCT.
% \revision{ Whereas a geolocation-based DCT system could still be consulted by the health authorities to augment their manual contact tracing that also helps Bob.}

% After the release of TraceTogether, researchers in other areas develop similar systems for their country \cite{} or improve it with more security features \cite{}. For example, Bell et al \cite{bell2020tracesecure} proposed a BLE-based protocol TraceSecure, in which homomorphic encryption is used. Homomorphic encryption supports computation on encrypted data, by which exposure risks can be computed securely on the server-side but can only be decrypted by users themselves. In this way, exposure risks are only revealed to the users themselves and not revealed to the central server. However, because of the 4-bits status indicator used in the system, Bell’s method is not able to find contacts of patients before the patients have been medically diagnosed. 

Third, these schemes can only capture cases where the disease spreads through direct face-to-face contact. However, time-lagged indirect spreading situations such as spreading through contaminated surfaces or lingering aerosols in poorly ventilated spaces 
%airborne aerosols
cannot be captured by these schemes.

\subsection{Geolocation-based Contact Tracing}
% The basic idea of a geolocation-based DCT system is to first record each user’s past geo-trajectories based on GNSS systems' output. 
A geolocation-based DCT system records each user’s past geo-trajectories based on GNSS systems' output. 
The users’ exposure risks are calculated based on the number of intersections between users’ recorded trajectories and patients’ trajectories in the past. However, logging users’ trajectory information in a DCT system without any privacy protection must be avoided, as adversaries can easily infer users’ personal information, such as identity, home and work addresses, and daily activities, from the data. 

Designing a privacy-preserving geolocation-based DCT system is still a challenging problem, and only a few solutions 
% \revision{were}% are
% proposed that 
make efforts to protect users’ privacy. Kato et al's work\cite{kato2020pcttee} attempts to address the privacy issue by proposing a hardware-supported trusted execution environment in the server, where plaintext comparisons of trajectories are carried out. This scheme essentially gives the server full trust for both healthy users and patients' private data.  Another set of projects~\cite{berke2020assessing, OpenMined2020, reichert2020privacy,fitzsimons2020note} relies on crypto-protocols such as Private Set Intersection (PSI) or MPC to protect the trajectory privacy of healthy users from an untrusted server. These schemes allow two parties to find the intersection of their data without revealing data not contained in the intersection. Yes, these schemes still have limitations. First, these schemes have to give the server full trust to store patient trajectory data in unprotected form, essentially sacrificing patients’ privacy in exchange for healthy user's privacy. Unfortunately, survey data has shown that the general public is distrustful of such a DCT design~\cite{TimbergPrivacyConcern2020}.
%does not have such trust to servers and are reluctant to adopt DCT with such kind of design~\cite{TimbergPrivacyConcern2020}. 
Second, these schemes reveal to a healthy user the location and time of a contact with a diagnosed patient. Since the healthy user may remember who he/she met at the time and location, the patient's identity is essentially revealed by the location/time information. Third, similar to Bluetooth-based DCT, the privacy protection techniques used in these schemes also make them incapable of capturing indirect spreading of the disease.

Our proposed method seeks to protect the users’ privacy of both sick and healthy users from both untrusted servers and other users. In addition, our scheme seeks to support considerations of both direct and indirect spreading of COVID-19. More importantly, our proposed method is able to cooperate with Bluetooth technology without sacrificing privacy protection, thus addressing the GNSS's lack of indoor localization.
To the best of our knowledge, % As far as we know, 
we are the first privacy-preserving DCT protocol
to combine GNSS with Bluetooth technology.% which can combine the GNSS with the Bluetooth together. 

\section{System Overview}\label{sec:sys-overview}

\begin{figure}[th]
    \centering
    \includegraphics[width=1.0\columnwidth]{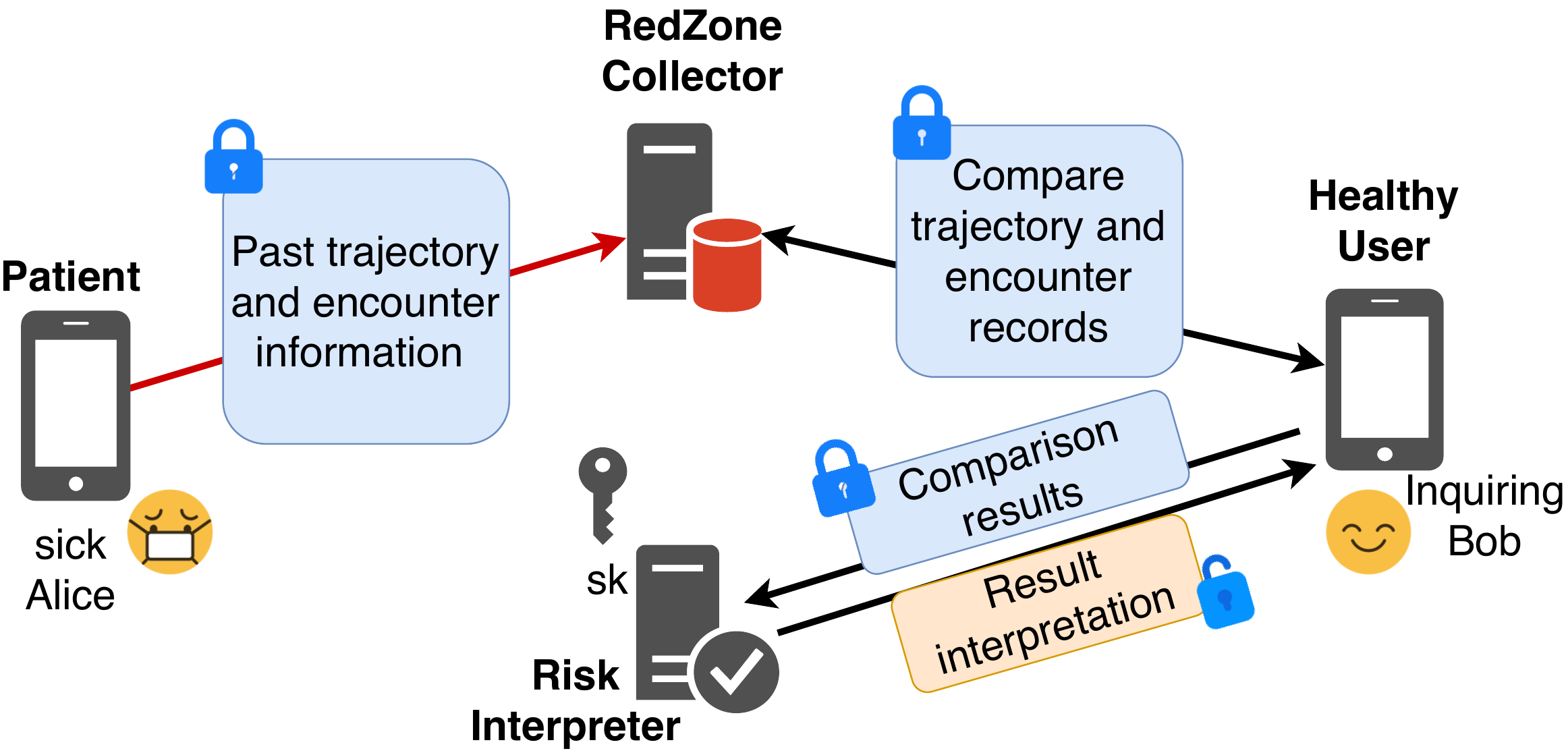}
    \caption{An overview of \sysName System. Patients, healthy user, \redzone and \riskInter are all involved in the MPC.}
    \label{fig:overview}
\end{figure}

\begin{table}[t]
    % \resizebox{1.0\columnwidth}{!}{
      \caption{Terms and definition.}
    \centering
    % \ra{1.3}
    \footnotesize
    \begin{tabular}{|c|p{6.2cm}|}
        
        \hline
        \textbf{Term} & \textbf{Definition} \\
        \hline
        \textit{Healthy user} &A healthy user of our contact tracing app. A healthy user may use our app to  check his/her exposure risk. \\
        \hline
        \textit{Patients} & Persons who have been diagnosed with COVID-19 and will share their contact tracing history to the DCT system. \\
        \hline

        % \textit{App user} & \begin{itemize}
        %     \item Uninfected User
        %     \item Patient 
        % \end{itemize} \\
        \hline 
        
        \textit{Contact} &  A person that has been in close proximity with a patient sometime during the past 14 days of incubation period. \cite{reichert2020survey}. \\
        \hline 
        
        \textit{Exposure risk} &Exposure risk reflects the probability that a person has been exposed to the virus. The more contact with patients, the higher the exposure risk. \\
        \hline

        \end{tabular}
    % \vspace{-0.8in}
    % }
  
    \label{tab:my_term}
    \vspace{-0.2in}
\end{table}

% \begin{enumerate}
%     \item \textit{Patients}: Persons who have been medically diagnosed with COVID-19 and use our contact tracing app. Patients decide to report their past trajectories as well as encounter information to the RedZone Collector after they got diagnosed. 
%     \item \textit{Contact}: A person that has been in close proximity with a patient sometime during the past 14 days of incubation period. \cite{cryptoeprint:2020:672}. 
%     \item \textit{Exposure risk}: Exposure risk level reflects the probability that a person has been exposed to the virus. The more past contacts that a person has with patients, the higher the risk level is. 
    
%     \item \textit{Normal user}: A user of our contact tracing app, who is not diagnosed with COVID-19. A normal user may use our app to  check his/her exposure risk level to the virus.
% \end{enumerate}

The design goal of \sysName is a hybrid DCT system that can provide accurate privacy-preserving exposure notification service to app users by leveraging both geolocation and Bluetooth technology. 
In this section, we introduce the threat model and briefly \revisionok{describe the workings and interconnection among \sysName components}.
% overview how each component of   \sysName works and how they cooperate with each other. 

%\subsection{System Model}

% yyang8: Does not seem to fit the bluetooth side of the system interaction

% A typical scenario of the system is described as follows. Firstly, patients update DCT servers with their contact tracing history, such as trajectory histories and broadcasted EphIDs. Then, any normal users who want to check their exposure risk will send the DCT servers a request for exposure risk computation along with their contact tracing history. Based on some epidemiological models, the server will compute the exposure risk by checking normal users' and patients’ history together, then the computation result will be returned to the normal users' app for further processing and an alarm will be raised if the exposure risk exceeds a risk threshold. 

% \begin{figure}[ht]
%     \centering
%     \includegraphics[scale=0.5]{fig/System_Model.pdf}
%     \caption{The system model of DCT }
%     \label{fig:sys_model}
% \end{figure}

\subsection{Threat Model}
Like all DCT systems, \sysName system also involves three types of interacting parties: 
patients, DCT servers, and healthy users. 
The terms ''patients" and ''healthy users" are defined in Table~\ref{tab:my_term}. 
DCT Servers are responsible for data storage and system operations. 
We consider two types of attackers against such a system:

{\em Adversarial users:} The goal of adversarial users is to use any possible strategies to deviate from the DCT protocol, and they can be either active or passive adversarial users. Active adversarial users actively participate in the protocol by injecting malicious messages. For example, an active adversary can fake as a patient and send spurious information to the central server or other users. Passive adversarial users refer to curious users who are interested in identifying other users' personal information, such as home addresses, workplaces, etc., by observing legitimate messages.  \revisionok{We assume that hardware is trusted in our system, and consider hardware attacks to be out of scope of the paper }. 
    
{\em Curious Servers:} A curious server refers to the honest-but-curious central server in a DCT system, usually maintained by tech companies, health authority or government. The objective of a curious server is to derive sensitive user information based on received protocol messages while following the DCT protocols. For example, the central server may learn users' home addresses, exposure risks, or build social graphs from the uploaded location information. If there are multiple  servers in the system, we assume that the servers do not collude with each other (e.g. operated by different agencies).
% \paragraph{Passive observers} may listen to all communication between app users and servers. For example, a passive observer may try to differentiate between positive and negative exposure risk using traffic analysis.  
% We assume that there is an authenticated and encrypted communication channel between each client and server, for example, using TLS~\cite{}. 
\revisionok{\subsection{\sysName Architecture}}

Figure~\ref{fig:overview} illustrates the architecture of \sysName, which includes patients, healthy users and two servers: \redzone and \riskInter.
% the , \sysName's design involves 
% \revisionok{two servers - (3) \redzone, and (4) \riskInter, apart from the (1) patients and (2) healthy users. }
% four parties: (1) patients, (2) normal users, (3) \redzone, and (4) \riskInter. 
\riskInter is a central server responsible for key \revisionok{distribution} and ciphertext decryption. \revisionok{The server } %\riskInter
generates a Paillier public/private key pairs $(pk,sk)$.  The public key, $pk$, is distributed to all users and the \redzone. The private key, $sk$, is kept as a secret only known to the \riskInter, which means \riskInter is the only party who can decrypt a Paillier ciphertext in the system. \redzone stores data and conducts contact tracing history comparisons between healthy users and patients. Both patients and healthy users have downloaded the \sysName smartphone app that locally logs a user’s trajectory and its encounter with other \sysName phones in the past two weeks, where the two-week time span is the maximum incubation period. 

\revisionok{When }% Once
a  user is tested positive with COVID-19 and becomes a patient, he/she can voluntarily let the app upload his/her past trajectories and contact event information to \redzone. 
The uploaded information is under strict privacy protection as it is masked using k-anonymity and  Paillier encryption before uploading. Manual contact tracing data about patients’ whereabouts can also be uploaded to \redzone by healthcare authorities using the same privacy-preserving mechanism. A healthy user, who wants to evaluate his risk of contracting the disease from 
\revisionok{patients, queries the \redzone to compute his exposure risk. 
% His trajectories are privately compared by the \redzone whereas the encounter records are privately compared by the smartphone.
\redzone performs secure computation to produce a  homomorphically encrypted query response. }
% , can query \redzone to privately compare his own phone’s trajectory and encounter records with the masked patient trajectory and encounter data from \redzone. The private comparison results are in homomorphically encrypted and their values are never revealed to \redzone.
\revisionok{The healthy user sends a masked version of the encrypted response to the \riskInter, which decrypts the masked results, and returns the decryption results to the healthy user. The healthy user removes the mask to turn the decryption results into the exposure risk.
}
% The normal user who initiates the private comparison obtains the results in encrypted form and sends the results to \riskInter, which decrypts the comparison results, converts it to an exposure risk metric to further improve privacy protection to patients, and returns the metric to the normal user.
The above \sysName operations are carried out under strong privacy protection, so that both  patients' and healthy users' trajectories and other private information are not exposed to app users, the servers or any other parties. (See Section \ref{section:geo-method} and \ref{sec:BLE-method} for details).

\section{Geolocation-based Contact Tracing} \label{section:geo-method}

\begin{figure*}[ht]
    \centering
    \includegraphics[width=0.95 \textwidth]{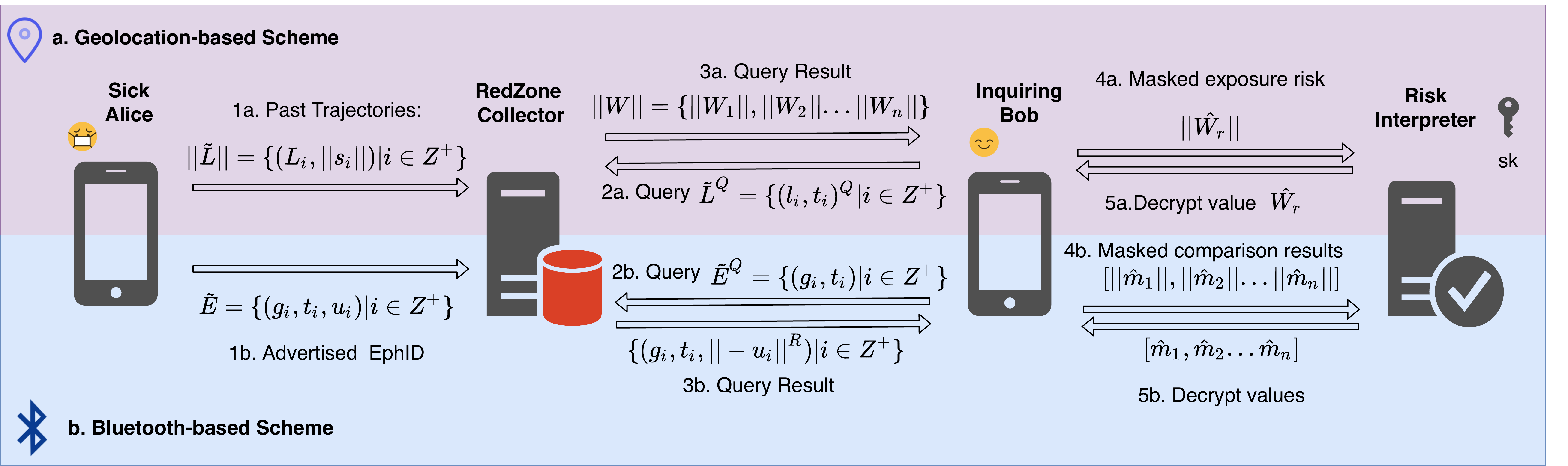}
    \caption{The data flows in \sysName for a). geolocation-based scheme and b). Bluetooth-based scheme.  }
    \label{fig:ble-based}
\end{figure*}

% The proposed \sysName system supports both GNSS and Bluetooth tracing while not sacrifices privacy.
This section depicts the details of our geolocation-based DCT system design. We first introduce the cryptosystem used in the design, and then discuss the three major design parts: patient data preparation, patient data integration, and exposure risk computation. 
% then we describe how the Bluetooth protocol is redesigned in our system to harmonize with geolocation-based protocol in the next section. 
Key notions are listed in Table~\ref{tab:my_symbol} in Appendix.

\subsection{Paillier Cryptosystem} 

\revisionok{\sysName leverages a partially homomorphic cryptosystem that allows homomorphic addition and scaling computations in the ciphertext domain.
}
% The \sysName will leverage a partially homomorphic cryptosystem, which allows addition, subtraction and scaling computations to be carried out directly in the ciphertext domain.
Since \sysName does not require fully homomorphic operations, its computational complexity is not high and can be scaled to handle large number of users.

We demonstrate a realization of the required homomorphic system by means of the Paillier Cryptosystem. The Paillier cryptosystem~\cite{paillier1999public} is based on arithmetic in the ring of integers modulo $N^2$, where $N$ is the product of two large primes. It is a probabilistic encryption mechanism that introduces a random value $\gamma$ in each encryption to ensure that encrypting the same message several times will, in general, yield different ciphertexts. It efficiently supports additive ($\oplus$) and scalar-multiplication  ($\otimes$) homomorphic operations. The details of Paillier cryptosystem’s operations are shown in Algorithm~\ref{alg:paillier} in Appendix, where the encryption result of a plaintext message \textit{m} is denoted by $\llbracket m \rrbracket$.

\subsection{Patient Data Preparation Procedure} \label{sec:geo_data_collect}

Suppose Alice is a patient and she installed the \sysName app before she gets infected. The app records Alice’s location trajectory \revisionok{and securely stores the trajectory on the phone~\cite{androidSecurityData, iosSecurityFile}. } The location trajectory is a list $\bm{L_r}:=\{ (l_i,t_i) | i \in \bm{Z}^+ \}$, where the tuple $(l_i,t_i)$ means that Alice visited location $l_i$  at time $t_i$. Location $l_i$ is defined by longitude and latitude. Once Alice is diagnosed with the disease, she can choose to voluntarily upload her trajectory data of the past two weeks to \redzone as follows. 

Alice's app first removes all locations that Alice has set to be sensitive (e.g. near home locations) from the recorded data. Then, the app synthesizes $k$ sets of fake trajectories $\bm{L_F} = \{\bm{L_{f1}},\bm{L_{f2}}...\bm{L_{fk}}\}$, where each fake trajectory $\bm{L_{fi}}$ needs to mimic real human trajectories. %Each entry of a fake trajectory $\bm{ L_{f_i}}$ is in the form of $\{(l_i,t_i)\}$. 
\revisionok{We also define a binary flag $s$ for each trajectory $\bm{L_{fi}}$ and $\bm{L_r}$. $s_r=1$ means trajectory  $\bm{L_r}$ is a real trajectory and $s_{fi}=0$ indicates  $L_{fi}$ is a fake trajectory. }
% We also define a binary flag $s_f$ for each trajectory $\bm{L_f}$. $s_f=1$ means trajectory  $\bm{L_f}$ is a real trajectory and $s_f=0$ indicates  $\bm{L_f}$ is a faked trajectories.  
The fake trajectories and real trajectories are mixed together to form a superlist  $\tilde{\bm{L}}$, where $\tilde{\bm{L}} = \{ (\bm{L_{f1}}, s_{f1}), (\bm{L_{f2}}, s_{f2}),..., (\bm{L_r}, s_r),...(\bm{L_{fk}}, s_{fk}) \} $, where the real trajectory $(\bm{L_r}, s_r)$'s position in the list is randomly picked.  In the next step, the flag associated with each trajectory is encrypted by Paillier cryptosystem using the public key, $pk$, that is published by the \riskInter. The encrypted flag for a trajectory $\bm{L_i}$ is denoted as $\llbracket s_i \rrbracket$ and the trajectory superlist containing encrypted flags is denoted as $\llbracket \tilde{\bm{L}}\rrbracket$, \revisionok{where $\llbracket \tilde{\bm{L}} \rrbracket:=\{	(\bm{L_i} ,\llbracket s_i \rrbracket ) | \forall \bm{L_i} \in \tilde{\bm{ L}}\}$.} Finally,  $\llbracket \tilde{\bm{L}}\rrbracket$ is uploaded to \redzone, \revisionok{as shown in step 1a in Figure~\ref{fig:ble-based}}.  Note that  \revisionok{$\llbracket  s_i \rrbracket$}, which flags Alice’s true and fake trajectory data, cannot be decrypted by \redzone since it does not have the decryption key.

Unlike Bluetooth-based method, the above Data Collection process also works with manual contact tracing. 
If Alice did not install the app before she is diagnosed, as long as she still remembers the places that she visited earlier, she can report these places to a contact tracing worker. The contact tracer can then manually convert this oral description into $\bm{L_r}$, use software to generate fake trajectories $\bm{L_F}$ and upload the true and fake trajectories to \redzone following the same data upload protocol as the app. By coordinating with manual contact tracing efficiently, \redzone will have a richer collection of patient data and hence discover more   encounters with patients. 

Note that before Alice uploads her trajectory information, an authentication process is required to ensure she is truly positive. \revisionok{The health authority provides a secret code in its positive test report to Alice, which is used for authentication and grants the app permission to upload data. The secret code remains valid only for a short period of time, to prevent selling the code to others.}

% such as using a secret verification code that comes with a positive test report or manual verification by health authorities. 

%Bluetooth-based methods always require patients have already installed the app before they got diagnosed. However, this is not the case in geolocation-based methods. 

% creates the encrypted trajectory data using the public key $pk$ as $\llbracket \bm{S} \rrbracket:=\{(\llbracket s_{lt} \rrbracket ,l,t)| \forall (l, t)\in \tilde{\bm{ L}}\}$, where $\llbracket s \rrbracket_{lt}$ is the encrypted flag indicates the reality of trajectory, that $ \llbracket s \rrbracket_{lt} =\llbracket1\rrbracket $,  if $(l,t) \in \bm{L}$ and $\llbracket s \rrbracket_{lt}=\llbracket 0 \rrbracket$ otherwise. The set $\tilde{L}$ serves as a cloak that hides the real trajectory of Alice in $\llbracket \bm{S} \rrbracket$. $\llbracket \bm{S} \rrbracket$ then is uploaded to RedZone Collector. Note that $\llbracket \bm{S} \rrbracket$, which holds Alice’s trajectory data, cannot be decrypted by RedZone Collector since it does not have the decryption key.

% \todo{add a more detailed description about the fake trajectory generation}

A major design trade-off in the above design is between communication cost and privacy protection. The number of fake trajectories (denoted as k) provides k-anonymity for the patient. The larger k is , the more secure the user's trajectory data will be. However, too large a k value  will also increase the amount of data to upload, process and store at \redzone.  Thus, the proper setting of k can be chosen according to network communication capacity and \redzone's process and storage capacity. 

In addition, to ensure real trajectories will not be exposed even if \redzone uses advanced data analysis tools, the fake trajectories need to have realistic mobility patterns as real trajectories do. We achieve this by adopting methods from~\cite{pappalardo2018data}, where each set of fake trajectories have different home addresses, workplaces and leisure places, and the time spent in those places remains similar to real human trajectories. Results from~\cite{pappalardo2018data} have shown that the method has a good capability to mimic the real trajectories. Other trajectory synthesis algorithms can always be used if they can imitate real trajectories well.

\subsection{Patient Data Integration }
% \paragraph{Data Integration:}
Although \redzone  cannot decrypt Alice’s trajectory data, it can still integrate Alice’s input with other patients’ data into a 2-dimensional  matrix $ \llbracket \bm{R} \rrbracket := \{\llbracket r \rrbracket_{lt}\}$ for all possible location $l$ in the service region and time point $t<2$ weeks. 
$\llbracket \bm{R} \rrbracket $ is an encrypted historical record of virus density at various (location, time) points, which are computed as follows. 

When initializing $\llbracket \bm{R} \rrbracket $'s  entry $\llbracket r \rrbracket_{lt}$,  set  $\llbracket r \rrbracket_{lt}\leftarrow \llbracket 0\rrbracket $ and mark the state of $\llbracket r \rrbracket_{lt}$ as "untainted".  For every uploaded patient trajectory $\bm{L_i}$ and every location $l \in \bm{L_i}$,  the accumulated amount of virus that the patient had expelled to location $l$ at  time $t$ can be expressed  by:
\begin{equation}
    \llbracket w_{lt} \rrbracket =
   \mathlarger \oplus_{\begin{subarray}{1}t\geq \tau\geq t-\delta\\
   (l, \tau) \in \bm{L_i}
   \end{subarray}} \quad \omega (t-\tau)\otimes \llbracket s_i \rrbracket,  
\end{equation}
Here, $\omega(t-\tau)$ is a positive decreasing weight function that captures the dying down of COVID-19 virus in an environment over time \cite{suman2020sustainability}, and $\delta$ represents the maximum lifespan of COVID-19 virus in a normal environment.  $\omega(.)$ is introduced to capture both direct person-to-person transfer cases and the cases where COVID-19 spreads indirectly through contaminated surface or aerosols. The exact shape of $\omega(.)$ and setting of $\delta$ are controlled by \redzone and can be updated according to new medical science discoveries or environment consideration at location $l$. 

The virus load $\llbracket w_{lt} \rrbracket $ is then integrated into $\llbracket R \rrbracket$. If  entry $\llbracket r \rrbracket_{lt}$'s state is "untainted", \redzone assign $ \llbracket r\rrbracket_{lt} \leftarrow \llbracket w_{lt} \rrbracket $ and mark  $\llbracket r\rrbracket_{lt}$'s state as "updated". If  $\llbracket r \rrbracket_{lt}$'s state is already "updated", 
\redzone computes $ \llbracket r\rrbracket_{lt} \leftarrow \llbracket r \rrbracket_{lt} \oplus \llbracket w_{lt} \rrbracket $.

After the aggregation,  if an entry $\llbracket r \rrbracket_{lt}$ is the ciphertext of a positive number (denoted as $\mu$), it means some   patient(s) were at location $l$ no more than $\delta$ time ago and the value $\mu$ represents the virus load that is left by these patients at location $l$ at $t$. $\llbracket r \rrbracket_{lt}=\llbracket 0 \rrbracket$ happens when no patient visited location $l$ recently.  %It can be easily seen that $ \llbracket \bm{R} \rrbracket$ essentially is the encrypted patient density map for the past 14 days.

To ensure  $ \llbracket \bm{R} \rrbracket$ only maintains history in the past 14 days, \redzone periodically updates $ \llbracket \bm{R} \rrbracket$ by deleting $\llbracket r\rrbracket_{lt}$ entries 
with $t$ earlier than 14 days ago.

\subsection{Exposure Risk Computation}~\label{subsec:geo-exp-risk-Q}
% \paragraph{Exposure Risk Query}
Now, consider a healthy user  Bob who wants to evaluate his risk of contracting COVID-19.  Using the same method of k-anonymity through fake trajectories , Bob’s app generates a trajectory superlist $\bm{\tilde{L}^Q}$, which includes both Bob’s real trajectory history $\bm{L_r^Q}$  and a $k'$ number of fake trajectories $\{\bm{L_f^Q}\}$.  As shown in Figure~\ref{fig:ble-based} step 2a, $\bm{\tilde{L}^Q}$ is sent to RedZone Collector as a query request.  For each trajectory  $ \bm{L_i^Q} \in \bm{\tilde{L}^Q}$ ,  \redzone calculates an encrypted exposure risk value $\llbracket W_i \rrbracket$ through homomorphic addition operations: 
\begin{equation}
    \llbracket W_i \rrbracket = \mathlarger \oplus_{\begin{subarray}{1}(l,t) \in \bm{L_i^Q}\\
    \text{state of $\llbracket r_{lt} \rrbracket$ is "updated"}
    \end{subarray}} \llbracket r_{lt} \rrbracket.
\end{equation}
% where $\llbracket W_{lt} \rrbracket $ stands for exposure risk at location $l$ at time $t$ and is calculated by 
% \begin{equation}
%     \llbracket W_{lt} \rrbracket =\mathlarger \oplus_{\begin{subarray}{1}t \leq \tau \leq t-\delta \\ \llbracket r\rrbracket_{l\tau} is "updated"
%     \end{subarray}}  \left (\omega (t-\tau)\otimes \llbracket r \rrbracket_{l\tau} \right ).
% \end{equation}
% Here, $\omega(t-\tau)$ is a positive decreasing weight function that captures the dying down of COVID-19 virus in an environment over time \cite{suman2020sustainability}, and $\delta$ represents the maximum lifespan of COVID-19 virus in a normal environment.  $\omega(.)$ is introduced to capture both direct person-to-person transfer cases and the cases where COVID-19 spreads indirectly through contaminated-surface-to-person transmission or aerosol transmission. The exact shape of $\omega(.)$ and setting of $\delta$ are controlled by \redzone and can be updated according to new medical science discoveries at anytime. 

% RedZone sends a list $ \llbracket U \rrbracket_{lt} := \{ \llbracket r \rrbracket_{l\tau} | t \leq  \tau \leq t-\delta \}$ back to Bob,  where $\delta$ represents the maximum lifespan of COVID-19 virus in a normal environment. 

Bob's app receives the list of encrypted exposure risks $ \llbracket \bm{W} \rrbracket = \{ \llbracket W_1 \rrbracket ,\llbracket W_2 \rrbracket,...,  \llbracket W_n \rrbracket \}$ from \redzone. Bob picks the one $\llbracket W_r \rrbracket $ that corresponds to his real trajectory since Bob knows the position of his true trajectory in the original $\tilde{\bm{L}}^Q$. Then, Bob needs to ask \riskInter to decrypt the exposure risk for him, which is the only entity in the entire \sysName system that holds the decryption key $(sk)$. In order to keep the exposure risk secret, Bob's app firstly generates a random masking number $\epsilon$ and then sends a masked exposure risk  $ \llbracket \hat{W_r} \rrbracket$ to \riskInter, where
\begin{equation}
    \llbracket \hat{W_r} \rrbracket := \llbracket W_r \rrbracket \oplus \llbracket \epsilon \rrbracket.
\end{equation}
\riskInter decrypts $\llbracket \hat{W_r} \rrbracket$ and returns the decryption result $\hat{W_r}$ to Bob's app. Bob's app then recovers his plaintext risk $W_r$ by removing the mask $\epsilon$ (a.k.a. $W_r = \hat{W_r} - \epsilon $) and display the exposure risk $W_r$ to Bob. $W_r\geq0$ indicates that there is a chance that Bob contracted COVID-19, either through a direct encounter with some patient(s) at some location(s) or visiting the location(s) a short time after the patient(s). The exposure risk is proportional to the virus load in the locations that Bob visited, which  considers the life span of COVID-19 virus in an environment. The larger the value of $W_r$, the greater the risk.  

%The app will query \redzone in a predefined frequency, such as once a day. 

% It is important to point out that the preliminary design aims at fast implementation and scalability in computation overhead. Stronger privacy protection can be achieved if we adopt more advanced private information retrieval, though the computation overhead can increase significantly. 

In the above design, both the patients' and the users'  real trajectories and the users' chance of contracting COVID-19 (a.k.a. exposure risk) are not revealed to any entity in the \sysName system except the user himself. % In the above design, none of Alice, Bob, and other patients and users’ real trajectories or chance of contracting COVID-19 are revealed to any entity in the \sysName system. 
Also, all the communications between clients and servers are through a secure TLS channel.
% Also, we can make all the communications between clients and servers happen through a secure TLS channel. 
Thus, users’ personal information cannot be leaked to any entity eavesdropping on the \sysName system. 
In addition, 
% Bob’s plaintext risk level is also only known to himself, which protect Bob's privacy from the servers and other users. Finally, 
the exposure risk is a single value, which does not reveal when and where encounters with patients happened. This ensures that Bob is unlikely to derive the identity of patients by cross-checking the exposure risk with his memory.

 \section{Bluetooth-based Contact Tracing} \label{sec:BLE-method}
%  To refine the contact tracing accuracy, \sysName also provides the Bluetooth option to users, if the app is allowed to use BLE, BLE beacons will be used for contact tracing.
 
 In this section, we introduce the 
 Bluetooth protocol of \sysName, which works with the geolocation-based protocol described in the previous section without compromising security and privacy. Besides Bluetooth, other technologies like ultrasound~\cite{loh2020flipping} can also be utilized for accurate close-contact detection and be combined with the geolocation side of \sysName. Without loss of generality, we choose Bluetooth to describe our design, since it is widely used in most existing DCT apps.

\subsection{Design Intuition}
% 1. BLE based limitations - to solve? server end comparison ? user end? problem on both side, so plain text is not okay, we need to use Paillier -> computation and communication overhead is huge? how to make it scalable? ID and location combination can make the computation range smaller -> another point, add location to BLE will make the system scalable and make it compatible with different regions

% 2. list design choices 
    % plaintext or crypto
    % ID and location assosication - scalable 
    % location region or more detailed location - will leak fake trajectory? -> need fake ID 
    
We start by describing the key design intuition behind our new Bluetooth-based protocol. 

% regardless whether they are centralized methods or decentralized methods, they all have privacy issues.
% Centralized methods may reveal users' exposure risks to a central server, which may not be trusted to hold such private data. Decentralized methods can protect exposure risks from being known by the central server. But it may leak patients’ identities to other querying users. This is because in an existing decentralized method, a healthy user compares its records of received EphIDs with patients' transmitted EphIDs in plaintext. Since the healthy user knows when an EphID was received, the comparison results reveal when the healthy user met with a patient. If the user happens to remember who he/she was with at the time, the patient's identity can be derived by the user. \todo{This paragraph is repetition of drawbacks from section 2.1, can consider to delete it to save some spaces.}

%Designing a Bluetooth-based method that mitigates these privacy issues simultaneously is a non-trivial task.

\paragraph{Design choice 1: Cryptosystem and MPC} Among the existing Bluetooth-based methods shown in Table~\ref{tab:related_works}, 
\revisionok{both centralized methods and decentralized methods have privacy issues. } We choose to solve the privacy problems of Bluetooth using the Paillier cryptosystem and MPC. Essentially, we can let the healthy users compare EphID records in ciphertext domain through homomorphic computation.   \riskInter then decrypts the encrypted comparison result and converts it into an exposure risk  that does not reveal patient encounter time. The exposure risk is then sent to the healthy user. By adding \riskInter to the comparison procedure, the encounter time with a patient is not revealed, thus protecting patients' identities from other curious users. 

Nevertheless, introducing the Paillier homomorphic cryptosystem raises the problem of communication overhead. In a typical existing decentralized system, healthy users usually download all the patients' EphID records from the central server for comparison. Since these records are downloaded in plaintext \revisionok{(e.g., 128-256 bits per plaintext)}, the data size is small and hence the scheme is scalable. However, in a privacy-perserving design, a healthy user needs to receive patients' EphIDs in ciphertext form (e.g., 2048 bits per ciphertext ) to ensure the patients' identities are not revealed to the healthy user.  This \revisionok{may} cause communication overhead if there is a large number of patient EphIDs.  To reduce the communication overhead, the number of records to be downloaded should be decreased,  which leads to our next design decision.  

\paragraph{Design choice 2: Integration of Spatio-temporal Information} Instead of pulling all patients' EphIDs from the central server, a healthy user can reduce communication overhead
\revisionok{by only retrieving EphIDs of patients whose reported trajectories (fake or real) intersect with the healthy user's trajectories. }
% if the user only retrieves a patient's EphIDs if the patient's trajectories intersect with the healthy user's trajectories. 
To achieve this, we decide to integrate location and time information with an EphID. Location information can be a region code representing a certain area, and time information can be the timestamp of broadcasting the EphID. As patients' EphIDs are uploaded, \redzone can organize EphIDs according to their locations and timestamps. Then, before downloading records, healthy users can specify the region and duration where trajectory intersections may happen and only retrieve EphIDs related to these trajectory intersections. 

In our design, we keep the granularity of region codes in EphIDs as a tunable parameter. The larger area a region code covers, the more encrypted patients'  EphIDs that the user needs to retrieve. While a smaller area can decrease the amount of data to download, it may also cause patients' past trajectories to be leaked. To understand this drawback, recall that our Bluetooth-based scheme should work with geolocation-based protocol together. When a patient uploads his advertised EphIDs containing region codes, his past trajectories are uploaded too. Then, a curious central server can distinguish fake trajectories if they are outside of the regions claimed by EphIDs. The smaller area a region code covers, the more fake trajectories will be exposed and removed by the central server. Thus, it is important to determine how to make the granularity of region codes small enough so that the communication cost can be reduced to a reasonable level while keeping users' trajectories protected. We solve this problem through the next design decision. 

% \paragraph{Location Granularity}

\paragraph{Design choice 3: Cooperation with fake trajectories} To ensure the EphIDs will not expose the fake/real status of uploaded trajectories, we make the uploaded patient EphIDs contain some fake EphIDs whose region codes are consistent with fake trajectories. The basic idea is that when a patient uploads advertised EphIDs, he not only claims that the EphIDs are broadcasted in the real visited places but also pretends EphIDs are broadcasted in other fake places. 

With the above design choices, our new Bluetooth-based protocol cooperates with geolocation-based protocol efficiently and privately. The main data flows in the protocol are demonstrated in Figure~\ref{fig:ble-based}, and the detailed procedures are described as follows.

\subsection{Ephemeral IDs generation}

%  \paragraph{Ephemeral IDs generation}
Once the Bluetooth option is turned on in a \sysName app, it  starts to generate EphIDs which are used to exchange with nearby \sysName devices. In our design, instead of using a random string~\cite{googleAppleDCT,troncoso2020decentralized}, an EphID is generated by concatenating the smartphone’s last available location with a random string. The detailed procedure for generating and exchanging EphIDs are described as follows: 

\begin{enumerate}[leftmargin=5pt]

    \item The first step is to generate unique and random strings  at each mobile device, using a similar method as existing Bluetooth methods~\cite{googleAppleDCT,troncoso2020decentralized}. Specifically, to avoid location tracking via the EphIDs, the random string used in EphID changes at every minute. We denote the sequence of these random strings as $\{u_i\}$.
    % At the beginning of each day, \sysName app generates a $seed$ and uses the seed to generate a set of random strings $\{u_i\}=
        % u_1 || u_2 || ... || u_n = PRG(seed),$
    % where PRG(seed) is a pseudorandom generator (e.g. AES in counter mode) which takes $seed$ and produces $n * 16$ bytes. 
    
    \item In the second step, \sysName app reads the last available location and converts it to an $n-character$ geohash-code. Geohash is a public domain function that encodes a geographic location into a short length of geohash code containing letters or digits, and each geohash code stands for a grid on the map where shorter codes mean larger grids. For example, an $8-character$ geohash code stands for a $19m\times 19m$ grid on the map. We denote geohash code as $g_i$.

    % \item The timestamp $t_i$ of broadcasting the EphID is also recorded, the resolution of the timestamp is 1 minute. 
    
    % \item \sysName read Mobile Network Code(MNC), Cell Identifier(CI) and  assigned cellular host IP address from the device, then the last $n$ bits (eg. $n = 8$) are extracted from the IP address. The value of extracted bits are concatenated with the MNC \& CI, the concatenated cellular information is denoted as $cell_i$.

    \item In the third step, the geohash code $g_i$ and a random string $u_i$ are concatenated into a string, and this string is the EphID $e_i = g_i ||  u_i$, where $||$ is a concatenation operator. For every minute, a new EphID is generated.
    
    \item In the final step, once the EphID is generated, it is continuously broadcasted by the app. All the used EphIDs along with the broadcast timestamps (1-minute granularity)  in 14 days, denoted as $\bm{E_r}:=\{(g_i,u_i,t_i) | i\in Z^+\}$ and called temporal-EphIDs,  will be stored securely on the phone.  
    \item When the app receives another device's EphID $e_i=g_i||u_i$, it will first check if the geohash code $g_i$ is within $\phi$ meters from itself. If the answer is yes, a temporal-EphIDs $(g_i,u_i,t_i)$ will be created for the EphID and be saved on the phone in the received EphID set $\bm{E_r^{Q}}$. Otherwise, the received EphID will be discarded. The purpose of comparing the geohash code $g_i$ in the received EphIDs with the receiver's location is to avoid Bluetooth relay and replay attacks, which are two of the main weakness in existing Bluetooth DCT works~\cite{reichert2020survey,gvili2020security,Vaudenay2020AnalysisOD}. In these two attacks, attackers collect EphIDs transmitted by a patient and replay the EphIDs in other locations to create false contact cases. As long as these other locations are more than $\phi$ meters away from the patient's location, such replayed EphIDs will be dropped by our scheme's comparison of the geohash code in the EphIDs with receiver's location. More discussions regarding the relay attacks can be found in Appendix~\ref{sec:supp_secu_analysis}. 
\end{enumerate}

% For example, Alice’s \sysName App read its location with the latitude of $37.2316625$ and longitude of $-80.425799$, then the geocode of this location is calculated as “DNWGCCGD”. In addition, Alice’s smartphone’s cellular IP address is 100.105.82.35 with a MNC 480 and cell id 53349424, and the last 8 bits of the IP address are saved as 35. After that, the app will generate EphID by concatenating the geocode, timestamp, and IP bits in the order of $location-MNC-CI-IP$, which is “DNWGCCGD-480-53349424-35”. By inlcuding geo-location, MNC, CI and IP address, each device’s EphIDs are ensuring to be unique at that time. 

% Once the EphID is generated, it will be continuously broadcast by the app, and record others’ advertised EphIDs, both broadcast and received EphIDs will be stored locally on the smartphone. Besides, The timestamp will also be recorded by the app while broadcasting / receiving a new EphID, which is denoted as $t_i$. 

% The reason for checking geohash code from received EphIDs is to avoid any Bluetooth relay attacks which are concerned as a serious security vulnerability in existed works~\cite{cryptoeprint:2020:672,gvili2020security,Vaudenay2020AnalysisOD}.

\subsection{Infected User Operation}
% \paragraph{Infected User Operation:} 

\begin{figure}[h]
    \centering
    \includegraphics[width=0.95 \columnwidth]{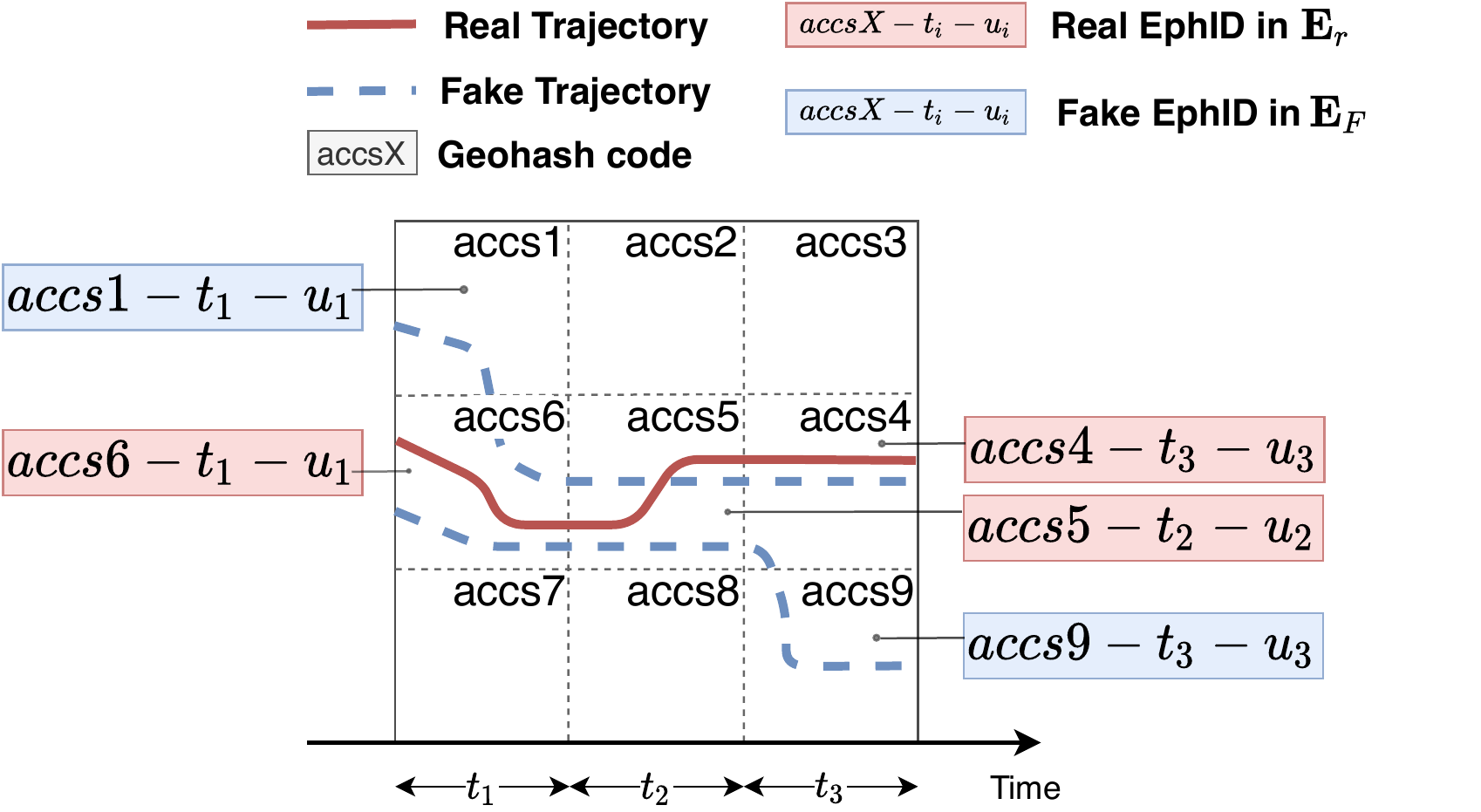}
    \caption{Demonstration of real and fake EphIDs. The map is divided into grids and represented by geohash codes. \sysName app generates fake and real EphIDs according to real and fake trajectories spanning $[t_1,t_2,t_3]$.}
    \label{fig:ephid}
\end{figure}

Once Alice has been diagnosed with the disease, she can choose to upload  $ \bm{E_r} = \{(g_i,u_i, t_i)| i \in Z^+\}$, which includes her advertised BLE EphIDs along with timestamps, to \redzone.  Before Alice sends her EphIDs, to prevent the EphIDs to compromise the privacy of the geolocation-based side of \sysName, a cloaking process is used to generate a superset $\bm{\tilde{E}}$ of $\bm{E_r}$, where $\bm{\tilde{E}} = \bm{E_r} \cup \bm{E_F}$. Here, $\bm{E_F}$ is a set of fake temporal-EphIDs which are generated based on the fake trajectories in $\bm{L_F}$ as shown in Figure \ref{fig:ephid}. First, all the  $(g_i,t_i)$ that 
have been visited by the fake trajectories are identified by converting $(l_i,t_i) \in \bm{L_f}\in \bm{L_F}$ to $(g_i,t_i)$.
If any $(g_i,t_i)$ does not appear in $ \bm{E_r}$, the corresponding $(g_i,u_i, t_i) $ is added into $\bm{E_F}$, where $u_i$ is the real $u_i$ broadcasted at $t_i$. Otherwise, the $(g_i,t_i)$ is discarded. Finally, the EphID superset $\bm{\tilde{E}}$ and encrypted location superset $\llbracket \tilde{\bm{L}}\rrbracket$ will be uploaded to RedZone Collector together. 

% \begin{equation}
% \bm{E_F} = \{(GEO(l_i),t_i,u_i) | \forall (l_i,t_i) \in \bm{L_F} and (GEO(l_i),t_i) \not\in )  \}
% \end{equation}

% The $GEO(.)$ above is the geohash function and $u_i$ is generated using the same pseudorandom function in Equation~\ref{equ:random} but with different seeds. 

% The fake EphIDs in $B’$ are consistent with fake trajectories in $L’$, each fake location $(l’_i, t’_i ) \in L’$ will be used to synthesize a corresponding fake EphID $b'_i = l’_i \mdoubleplus cell’_i $ in that time $t'_i$, where $cell’_i$ is a random value. After that, the $cell_i$ in each EphID is homomorphically encrypted, and the encrypted EphID is denoted as $\llbracket b_i  \rrbracket =l_i \mdoubleplus \llbracket cell_i  \rrbracket $. 

% as showed in step \ding{173} in Figure \ref{fig:ble-based}.

\subsection{Data Integration}
% \paragraph{RedZone Collector:} 

Once \redzone received the temporal-EphIDs from patients, it will first extract $u_i$ from the EphIDs, reverse its sign and encrypt $-u_i$  by Paillier cryptosystem. It stores the ciphertext $\llbracket -u_i \rrbracket$ and indexes the storage location using the geohash code $g_i$ and timestamp $t_i$. \redzone does not know which EphID is real or fake because their geohash codes are consistent with locations in $\llbracket \tilde{\bm{L}}\rrbracket$, and all the $u_i$ has the same pattern as they are generated using the same pseudo-random function. \redzone will automatically delete EphIDs older than 14-days. 
 
\subsection{Healthy User Query} 

As shown step 2b in Figure~\ref{fig:ble-based}, when a healthy user Bob wants to check his exposure risk, his \sysName app queries \redzone by uploading  superset  $\bm{\tilde{E}^{Q}} =\{ (g_i,t_i) | i \in Z^+\}= \bm{E_r^{Q}} \cup \bm{E_F^{Q}}$. Here, $\bm{E_r^{Q}}$ holds true $(g_i, t_i)$ from the geohash codes of the EphIDs
% received temporal-EphID set $\bm{E_r^{Q}}$,which holds the EphID 
that Bob received in the last 14 days with timestamps of the reception.  $\bm{E_F^{Q}}$ holds fake $(g_i,t_i)$ pairs that are generated based on the set of fake trajectories $\{\bm{L_f^Q}\}$ at the geolocation side protocol (see section \ref{subsec:geo-exp-risk-Q}).  These fake $(g_i,t_i)$ pairs ensure the true location trajectory will not be revealed.  Also, to reduce the false-positive rate, only EphIDs with sufficient contact duration will be selected for the query, those EphIDs with contact duration under the threshold will be ignored.

%Finally, fake EphIDs $\bm{E_F^{Q}}$ and real received EphIDs $\bm{E_r^{Q}}$ forms a superset $\bm{\tilde{E}^{Q}} = \bm{E_r^{Q}} \cup \bm{E_F^{Q}} $  and uploaded to RedZone Collector for exposure risk query. 

% The difference is that instead of using the Bob's fake locations themselves, locations in fake received EphIDs should be around Bob's fake locations. The insight is that locations of EphID advertiser can not always share the exact same positions with the recipient, it can show up in the nearby places, otherwise, the EphIDs will look suspicious. 

% \note{archanaa: In this paragraph, it is not clear which $u_i$ is plaintext and which is ciphertext}
Once \redzone receives the query request, RedZone Collector retrieves EphIDs that matches $(g_i, t_i)\in \bm{\tilde{E}^{Q}}$ in its own storage. If there is a match, \redzone sends the matching EphID's $u_i$, denoted as $u_i^R$, back to Bob in the encrypted form of $(g_i,t_i,\llbracket -u_i^R \rrbracket)$. After Bob's app get all the responses from \redzone, it discards the $\llbracket -u_i^R \rrbracket$ corresponding to the fake $(g_i, t_i)$. Then, it performs the homomorphic addition between the rest of $-u_i^R$ and the $u_i$ from locally stored received EphIDs: $ \llbracket m_i  \rrbracket =  \llbracket u_i  \rrbracket \oplus  \llbracket -u_i^R \rrbracket$, where  $m=0$ if $u_i = u_i^R$, which means Bob have a contact with an infected user, otherwise, $m \neq 0$. After checking all the returned results, Bob will get a list of encrypted subtraction results $\llbracket \bm{M}  \rrbracket = \{\llbracket m_1  \rrbracket, \llbracket m_2  \rrbracket ... \llbracket m_n  \rrbracket\}$.

% RedZone Collector will do a homomorphic subtraction between the infected user's EphID and the query EphID: $ \llbracket m  \rrbracket =  \llbracket cell^{patient}  \rrbracket -  \llbracket cell^{query}  \rrbracket$, where $\llbracket m  \rrbracket$ is a encrypted subtraction result and  $m=0$ if $cell^{patient} = cell^{query}$, which means Bob have a contact with an infected user, otherwise, $m \neq 0$. For all matched geocodes and timestamps, RedZone Collector will get a list of encrypted subtraction results $\llbracket M  \rrbracket = \{\llbracket m_1  \rrbracket, \llbracket m_2  \rrbracket ... \llbracket m_n  \rrbracket\}$. After that, $\llbracket M  \rrbracket$ is returned to Bob's app. 

Note that Bob's concern is how many infected users he has contacted in the past 14 days, which is equivalent to how many $ \llbracket 0  \rrbracket\ $ exist in $\llbracket\bm{M}  \rrbracket$. To figure out that, as showed in step 4b and 5b in Figure \ref{fig:ble-based}, Bob's app sends a masked comparison result list $\llbracket \bm{\hat{ M }} \rrbracket = \llbracket \bm{M}  \rrbracket \oplus \llbracket \epsilon  \rrbracket$ to Risk Interpreter. Risk Interpreter decrypts  $\llbracket \bm{\hat{ M }} \rrbracket$, randomly shuffles the sequence of entries in the decrypted lists and returns the final list $\bm{\hat{M}} = \{ \hat{m_1}, \hat{m_2} ...\hat{m_n} \} $ back to Bob's app, which then recovers the $\bm{M}$ by removing the mask $\epsilon$ from all list entries. Now Bob can see how many infected users he has contacted, which is the number of 0 in the list $M$.

In the above design, the exposure risk is  only known to Bob himself. \redzone does not know because the EphID comparisons are conducted on the smartphone. \riskInter does not know because the comparison results  $\llbracket \bm{\hat{M}}  \rrbracket$ is masked by $\epsilon$. What’s more, Bob can not infer encounter time/location with patients because the shuffling of entries in $\hat{ M} $ before it is returned from \riskInter decouples the comparison results from location and time information. Besides, none of the healthy users' and infected users' trajectories, social graphs are revealed to any entity in or outside of the \sysName system because they are hidden under fake trajectories and fake EphIDs.

\section{System Security Analysis} \label{sec:sec_analysis}

\begin{table}[t]
\caption{Possible attacks and our defense mechanisms}
\centering
\resizebox{\columnwidth}{!}{%
\begin{tabular}{|l|l|}
\hline
\textbf{Attack}                         & \textbf{Our Defense Mechanisms }                                     \\ \hline
Cross-reference         & Fake EphIDs + fake trajectories.                              \\ \hline
One-contact/place       & \begin{tabular}[c]{@{}l@{}} User can remove sensitive records before \\ uploading them.  \end{tabular}             \\ \hline
 Relay and Replay        & \begin{tabular}[c]{@{}l@{}}Adding Geolocation and timestamp \\ information to EphIDs.                    \end{tabular} \\ \hline
Aggressive Broadcasting & \begin{tabular}[c]{@{}l@{}}Geolocation protocol can work without \\ Bluetooth \end{tabular}

 \\ \hline
Data Poisoning          & \begin{tabular}[c]{@{}l@{}} Patient authentication and secure data \\ storage.\end{tabular}                 \\ \hline
Man in the Middle      & Communicate through TLS channel.                              \\ \hline
\end{tabular}%
}

\label{tab:security_table}
\vspace{-0.2in}
\end{table}
In this section, we discuss the possible attacks on \sysName and how they are mitigated in our design.
Table~\ref{tab:security_table} summarizes the attacks and our defense mechanisms, of which we discuss two main attacks in this section and discuss the rest in Appendix~\ref{sec:supp_secu_analysis}.
% The attacks and defense mechanisms are summarized in Table~\ref{tab:security_table}. In this section we discuss two of the attacks and leave the discussion on other attacks in Appendix~\ref{sec:supp_secu_analysis}

\subsection{Cross-reference Attack}
\revisionok{A cross-reference attack occurs when an attacker attempts to get users' personal information by cross-checking different types of data. In our design, we consider \redzone as a possible attacker since it accepts data from both patients and healthy users. The goal of \redzone is to pick out an app user's real trajectories from the cloaking trajectories, it can either cross-check trajectories with the EphIDs from the same user or cross-check trajectories with other users' uploaded records. However, our protocols can prevent such attacks in both ways because of the following reasons.}

\revisionok{First, when the attacker cross-checks cloaking trajectories with the user's EphIDs, it compares the $(g_i, t_i)$ in EphIDs with every $(l_i,t_i)$ from trajectories, and any trajectories containing a $(l_i,t_i)$ without a corresponding $(g_i, t_i)$ existed suggests a fake trajectory. \sysName's Bluetooth protocol prevents this happening as fake EphIDs are synthesized along with fake trajectories. It ensures that both real and fake $(l_i,t_i)$ always have a corresponding $(g_i,t_i)$ in EphIDs.}

\revisionok{Second, when the attacker cross-checks a user's cloaking trajectories with others users' uploaded records, it checks whether the user's trajectories intersect with others' while their EphIDs exchanges at the same $(l,t)$. The insight is that intersections among real trajectories usually are accompanied by some exchange of EphIDs, while fake trajectories' intersections do not have such records of EphIDs exchange. However, in the proposed system \redzone cannot exam whether two users exchange EphIDs, since patients only upload their advertised EphIDs to \redzone, while healthy users downloads patients EphIDs and exam EphIDs exchange on the user-side. To sum up, the protocol design of \sysName can securely protect users' privacy from the cross-reference attack.}

\subsection{One-contact/place Attack}
 If an attacker wants to identify the infect status of an individual, the attacker can create a new account and only contact the person or stay at the place which is highly related to the individual (e.g. home or workplace). We call this one-contact/place attack. In \sysName, a patient has redacted specific locations/time before they upload their trajectories and EphIDs to the \redzone, which prevents this attack from happening. For example, an attacker who wants to know the infect status of his neighbor can deploy a one-place attack targeting the home/workplace of the neighbor. However, if the neighbor is infected but chooses to remove any trajectories and EphIDs associated with home/workplace before sharing records with the  \redzone, the attacker can will not get a high exposure risk regarding the targeted place.

%  patients to redact periods of time or places that they feel sensitive, then the trajectories and EphIDs related to redacted time/places will be deleted before uploading to \redzone. 

% \input{privacy_analysis}
% \input{security_analysis}

\section{Evaluation}
\label{sec:evaluation}
In this section, we demonstrate \sysName's interoperability, and its scalability to large deployments.

\begin{figure*}[t]
\centering
     \begin{subfigure}[t]{0.32\textwidth}
     \includegraphics[width=\columnwidth]{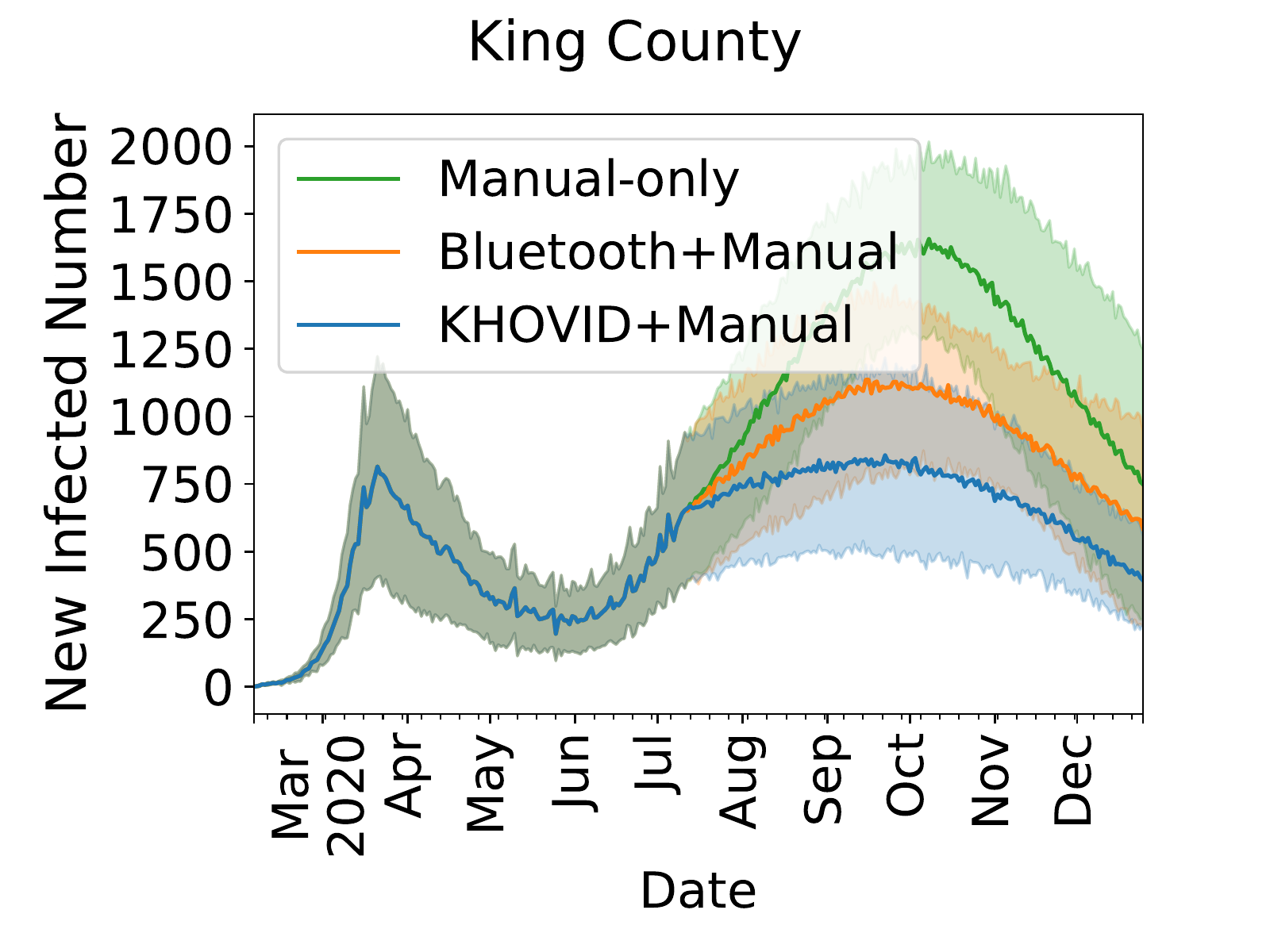}
     \label{}
     \end{subfigure}
     ~
    \begin{subfigure}[t]{0.32\textwidth}
     \includegraphics[width=\columnwidth]{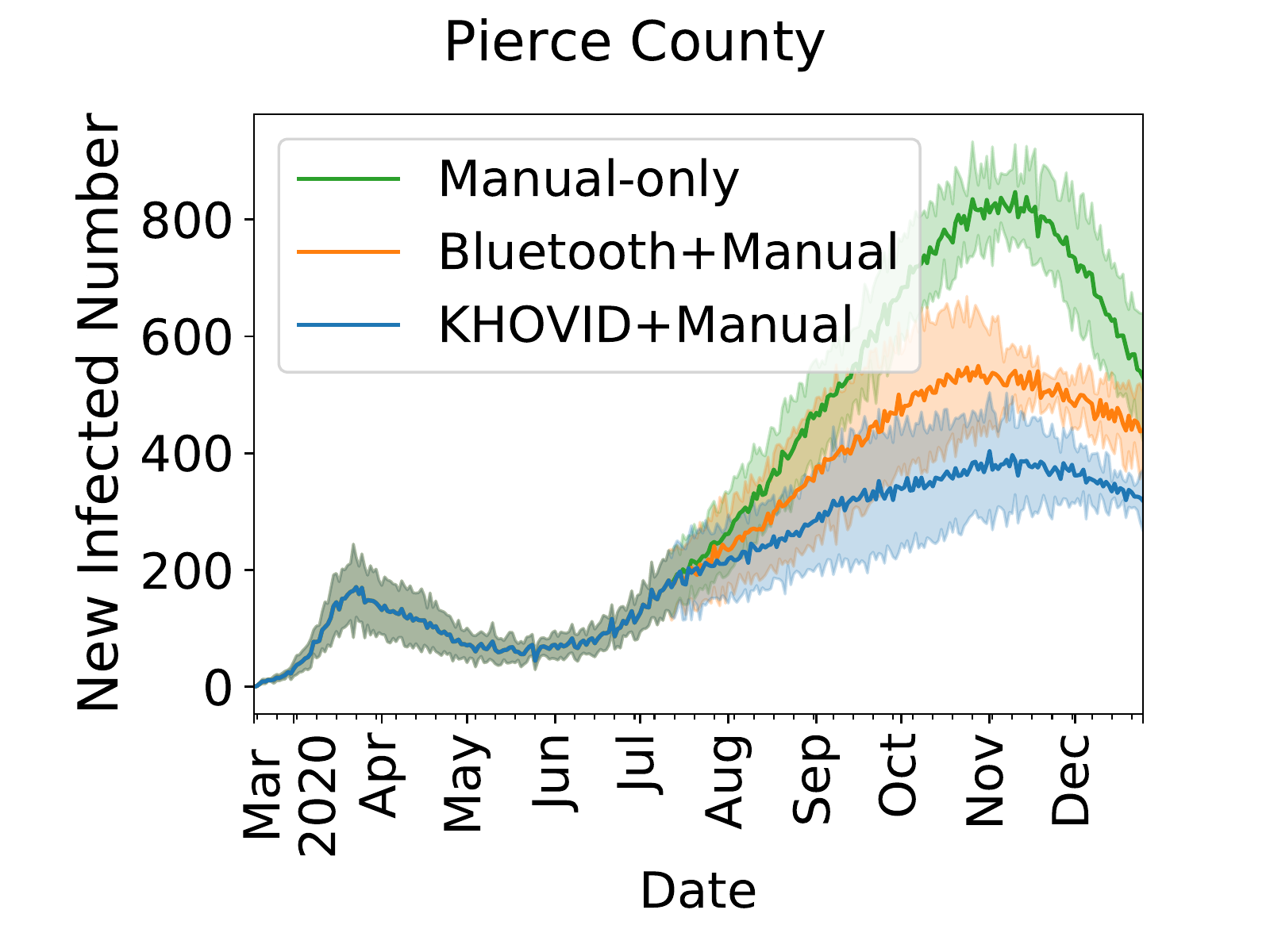}
     \label{}
     \end{subfigure}
     ~
     \begin{subfigure}[t]{0.32\textwidth}
     \includegraphics[width=\columnwidth]{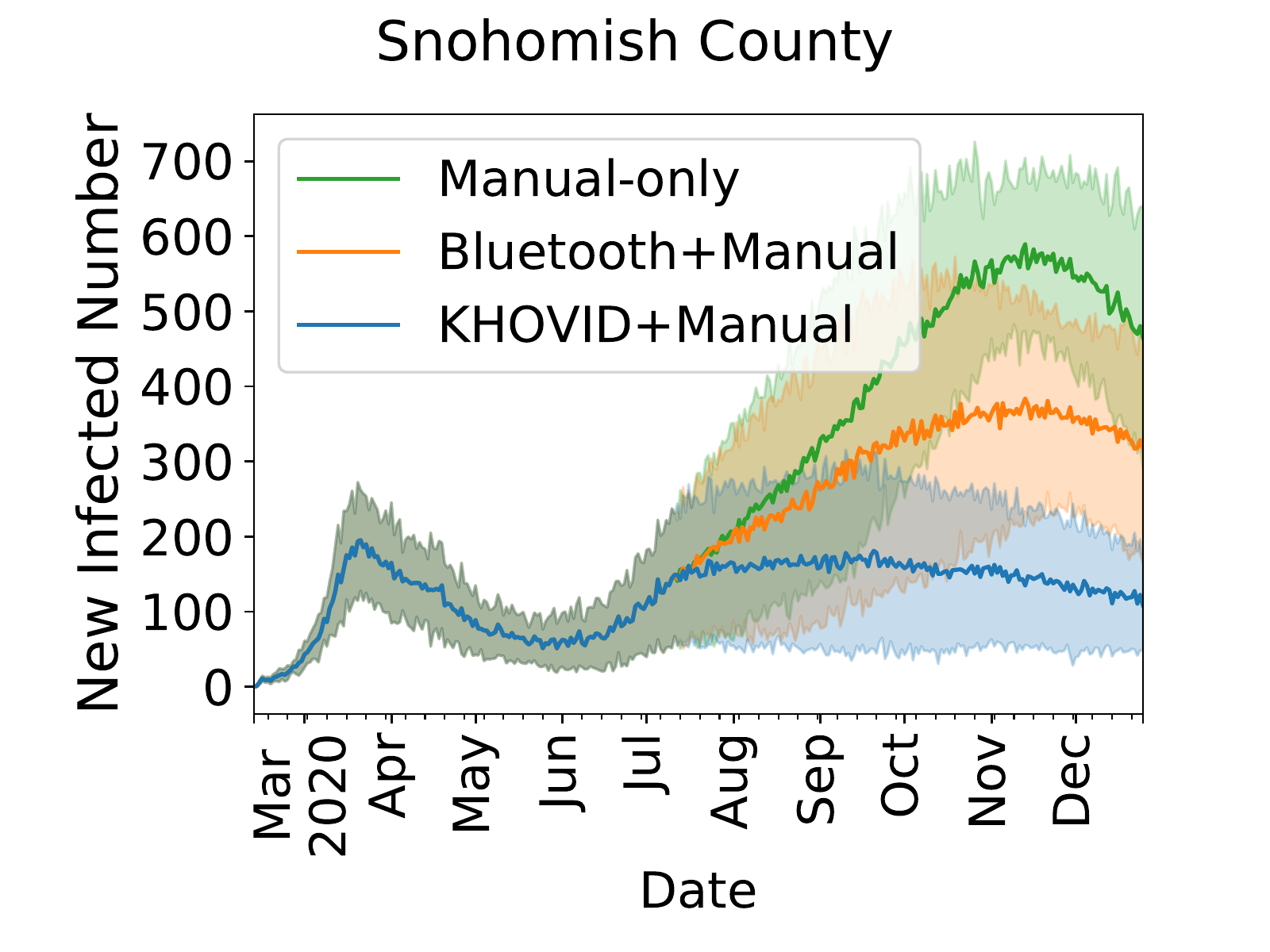}
     \label{}
     \end{subfigure}
    \vspace{-0.3in}
    \caption{Simulation result for daily new infected number in King county, Pierce County and Snohomish County under  Manual only tracing, Bluetooth-DCT + Manual tracing, and \sysName + Manual tracing. All DCTs are deployed on July 11. The shaded areas represent the 95\% confidence interval. }
    \label{fig:sim_adoption}
%\hfill%
\vspace{-0.14in}
\end{figure*}

\begin{table}[t]
\caption{Key parameter settings in the simulation of adoption rate}
\centering

\resizebox{\columnwidth}{!}{%
\begin{tabular}{|l|c|c|c|}
\hline
                                                                                               & \textbf{Manual} & \textbf{Bluetooth} & \textbf{\sysName} \\ \hline
DCT adoption rate                                                                                  & 0               & 0.3                & 0.3                  \\ \hline
\begin{tabular}[c]{@{}l@{}}Fraction of infected DCT users\\  who upload records to server\end{tabular}    & 0               & 0.6                & 0.6   \\ \hline
\begin{tabular}[c]{@{}l@{}}Fraction of patients interviewed by \\ manual tracer that share manual \\tracing data with DCT \end{tabular}    & 0               & 0                & 0.4   \\
\hline
\begin{tabular}[c]{@{}l@{}}\# of tracers per 100k people \end{tabular} & 15            & 15               & 15                 \\ \hline

\begin{tabular}[c]{@{}l@{}}\# of notification per day per contact \\ tracer \end{tabular}       & 3               & 3                  & 6                    \\ \hline

\end{tabular}%
}

\label{tab:sim_setting}
\vspace{-0.2in}
\end{table}

\subsection{Impact of \sysName's interoperability}
\label{sec:eval-adoption}

Interoperability, in this paper, means the ability to merge \sysName's DCT with manual contact tracing. Using a simulation of COVID-19 spreading, we compare \sysName with Bluetooth-based DCT. The simulation results show that by combination with manual contact tracing, \sysName can reduce infections more dramatically compared to Bluetooth-based DCT under the same adoption rate. 

%and geolocation-based DCT through simulations on an individual-based computational model. 

\subsubsection{Simulation Model}
\begin{sloppypar}
We ran the simulations based on an individual-based computational model called OpenABM-Covid19~\cite{hinch2020openabm}. OpenABM-Covid19 can evaluate non-pharmaceutical interventions during the Covid-19 pandemic, including both manual and digital contact tracing. It uses epidemiological and demographic parameters as a guide, seeking to simulate individuals and their interactions in home, work, and community contexts. OpenABM-Covid19 has been used to explore manual/digital contact tracing interventions in the UK ~\cite{hinch2020openabm}and Washington state in US~\cite{abueg2020modeling}. 
\end{sloppypar}

We set the pandemic model parameters in the simulations based on population, demographic, and occupational structure from King, Pierce, and Snohomish counties in Washington state. Three different tracing modes are considered: (1) {\em Manual only}, where contact tracing is carried out by hand, (2) {\em Bluetooth + Manual}, where the two tracing schemes work independently from each other, and (3) {\em \sysName + Manual}, where the two tracing schemes collaborate with each other. We assume the adoption rate of \sysName or Bluetooth DCT is all 30\%.  Other important parameters are listed in Table~\ref{tab:sim_setting}.

Some of the parameters in Table~\ref{tab:sim_setting} are set differently for Bluetooth DCT and \sysName DCT because \sysName can collaborate with manual  tracing.
 First, the manual contact tracers can help patients who did not install \sysName app to provide trajectories to \sysName. We assume 40\% of the trajectory data identified by these manual interviews are shared with \sysName. Second, a patient who installed \sysName app can give manual contact tracer his recorded trajectory so that the interview time can be greatly shortened comparing with oral description of past trajectory. Assume that a manual tracer spent 80\% of his time in interviews with patients and 20\% of time in sending notifications to the contacts of the patients. Assume 30\% of the interviewed patients have \sysName app and can hence shorten the interview time by 80\%. These assumptions lead to a saving of 20\% of a contact tracer's time, which we assumed is invested to notify more contacts of the patients and leads to doubling in the contacts that a tracer can notify per day. Other parameters not listed are the same among three settings. A complete list can be found in~\cite{abueg2020modeling}. Each setting is simulated for 5 iterations, with a different random seed in each iteration. 

%If DCT is used, we assume 30\% adoption rate of DCT. We assume that 60\% of infected DCT users choose to upload their data to the server of DCT system. The number of manual tracers per 100k people is 15.  Since Bluetooth DCT cannot aid manual tracing, the number of notifications per day per contact tracer and the fraction of encounters with strangers that manual contact tracer can identify are the same regardless of Bluetooth DCT. They are set as 3 and 0.05 respectively. Since \sysName can provide risk levels at various locations to manual contact tracing, the 

%\sysName can provide manual contact tracing virus density information at public places. For example, geolocation-based DCT can help to trace more random contacts, thus we increase the fraction of traced random contacts from 0.05 to 0.35. Besides, the manual tracers can notify close contacts faster with the help of DCT apps, thus the number of daily notifications per contact tracer is increased from 3 to 6. 

\subsubsection{Simulation Results}
The simulation results are shown in Figure~\ref{fig:sim_adoption}, in which the daily infected numbers in the three counties are presented. Both {\em Bluetooth + Manual} and {\em \sysName + Manual} can reduce the new infection number compared with {\em Manual only} method. The number of infections prevented by using a geolocation-based DCT app is higher than by using a Bluetooth-based DCT app. Using a geolocation-based DCT app, the infection peak is reduced by 53.6\%, 48.9\% and 91.7\% in King, Pierce, and Snohomish counties, respectively.

\subsection{Scalability }
\label{sec:eval-scalability}
% \begin{figure*}[t]
% \centering
%     %\setlength{\abovecaptionskip}{-0.1cm}
%      \begin{subfigure}[t]{0.3\textwidth}
%      \includegraphics[width=\columnwidth]{fig/Bar-commu-k-squre.pdf}
%      \caption{}
%      \label{}
%      \end{subfigure}
%      ~
%     \begin{subfigure}[t]{0.3\textwidth}
%      \includegraphics[width=\columnwidth]{fig/Bar-compu-k-square.pdf}
%      \caption{Spoofing}
%      \label{}
%      \end{subfigure}
%      ~
%      \begin{subfigure}[t]{0.3\textwidth}
%      \includegraphics[width=\columnwidth]{fig/Bar-battery-k.pdf}
%      \caption{Spoofing}
%      \label{}
%      \end{subfigure}
%     \vspace{-0.12in}
%     \caption{}
%     \label{}
% %\hfill%
% \vspace{-0.14in}
% \end{figure*}

In this section, we evaluate the scalability of a \sysName prototype from three perspectives: (1) communication overhead, (2) computation overhead, and (3) battery consumption. We also compare with other DCT schemes. 

% First one is the detection efficiency, where we consider the detection accuracy, false alarm rate, and how they contribute to prevent spreading. The second one is Computation overhead that we consider how much computational power is needed to perform the \sysName. The last part is communication overhead which measures the needed internet resources

\subsubsection{\sysName Prototype}
We built the \sysName system by implementing a contact tracing app, a \redzone server, and a \riskInter server. The system is used as a proof-of-concept for evaluation. 

\paragraph{Contact tracing app} The contact tracing app is built on Android platform, in which we implemented the geolocation and the Bluetooth protocols based on an open-source Safe-Path framework~\cite{safePath} \footnote{We only use Safe-Path framework for logging trajectories}. The app has been tested over multiple phone models, including  Xiaomi MIX2 (Android8.0, Snapdragon 835), Realme X(Android10.0, Snapdragon 710 ), and Google Pixel 3 (Android 9.0, Snapdragon 835). All the functions inside the app follow the protocol described in section~\ref{section:geo-method} and ~\ref{sec:BLE-method}. 

% In addition, every day the app sends an exposure risk query automatically and prefer the time when the smartphone is charged and connected to the WiFi,  and an API for uploading the user's records is also provided in case the user is infected later. 

\paragraph{Central Servers} We also built a \redzone server and a \riskInter server in the system. The data collection, data integration, and exposure risk computation in the servers are implemented through Firebase~\cite{firebase}, which provides a NoSQL cloud database for app developments. Note that the servers can also be built on other platforms. In our evaluation, we tested the servers on a Firebase emulator running on a Ubuntu machine with an AMD Ryzen 93900 12-core processor with 16GB memory.

% \subsection{Detection Accuracy}
% This section serves to present the detection efficiency of \sysName. However, due to the in-feasibility of a considerable scale of human test, we implement a network agent-based model~\cite{} to simulate the efficiency of our system. 

% \todo{Add brief introduction of simulation model, system settings, model parameters, and simulation result}
% \todo{Accuracy of closing contact detection GPS only vs GPS+Bluetooth}
\subsubsection{Evaluation Settings}
\begin{table}[t]
\caption{Parameters settings of scalability evaluation}
\centering

\resizebox{\columnwidth}{!}{%
\begin{tabular}{|l|l|}
\hline
Life span of an EphID's random string $(\sigma) $      & 20 minutes       \\ \hline
Character length of an EphID's geohash code               & 8                \\ \hline
Location logging frequency                          & every 15 seconds \\ \hline
App's daily logging duration & 10-16 hours \\ \hline 

Concerned person-to-person contact duration & 15 minutes \\ \hline
Concerned dirty-surface-to-person duration $(\delta)$ & 2 hours          \\ \hline
Average number of daily close contacts~\cite{danon2013social}         & 108            \\ \hline

\end{tabular}%

}

\label{tab:settings}
\vspace{-0.1in}
\end{table}

To evaluate the scalability of \sysName, we ran an experiment where a healthy user installs the app. The app collects the smartphone's location and records EphID exchanges when the user is out-of-home. The app sends an exposure risk query every day. Two weeks later, the user is tested positive and decides to share his records to \redzone. The important parameters of the experiment are presented in Table~\ref{tab:settings}. Specifically, we set the app's daily logging duration to a random value between 10 and 16 hours. For each trajectory uploaded by the healthy user during a query, the number of intersections with patients' records is set to $108*0.013*k = 1.40*k$, where  $k$ is the parameter in k-anonymity and the number of average daily contacts is 108 (derived from ~\cite{danon2013social}). The probability that a contact is a patient is 0.013, which is calculated by assuming a two-week window over the maximum daily new infection rate (2k/2.2M=0.0009) from King County in Figure~\ref{fig:sim_adoption}. 

% In addition, for each intersection point, we set the average number of EphIDs downloaded from \redzone to 10.

%Note that the number of intersections can vary in different demographics, as people's visited places can be concentrated in a small town but more distributed in a big city. 

\subsubsection{Measurement Results}

\begin{figure}[t]
    \centering
    \includegraphics[width=1.0 \columnwidth]{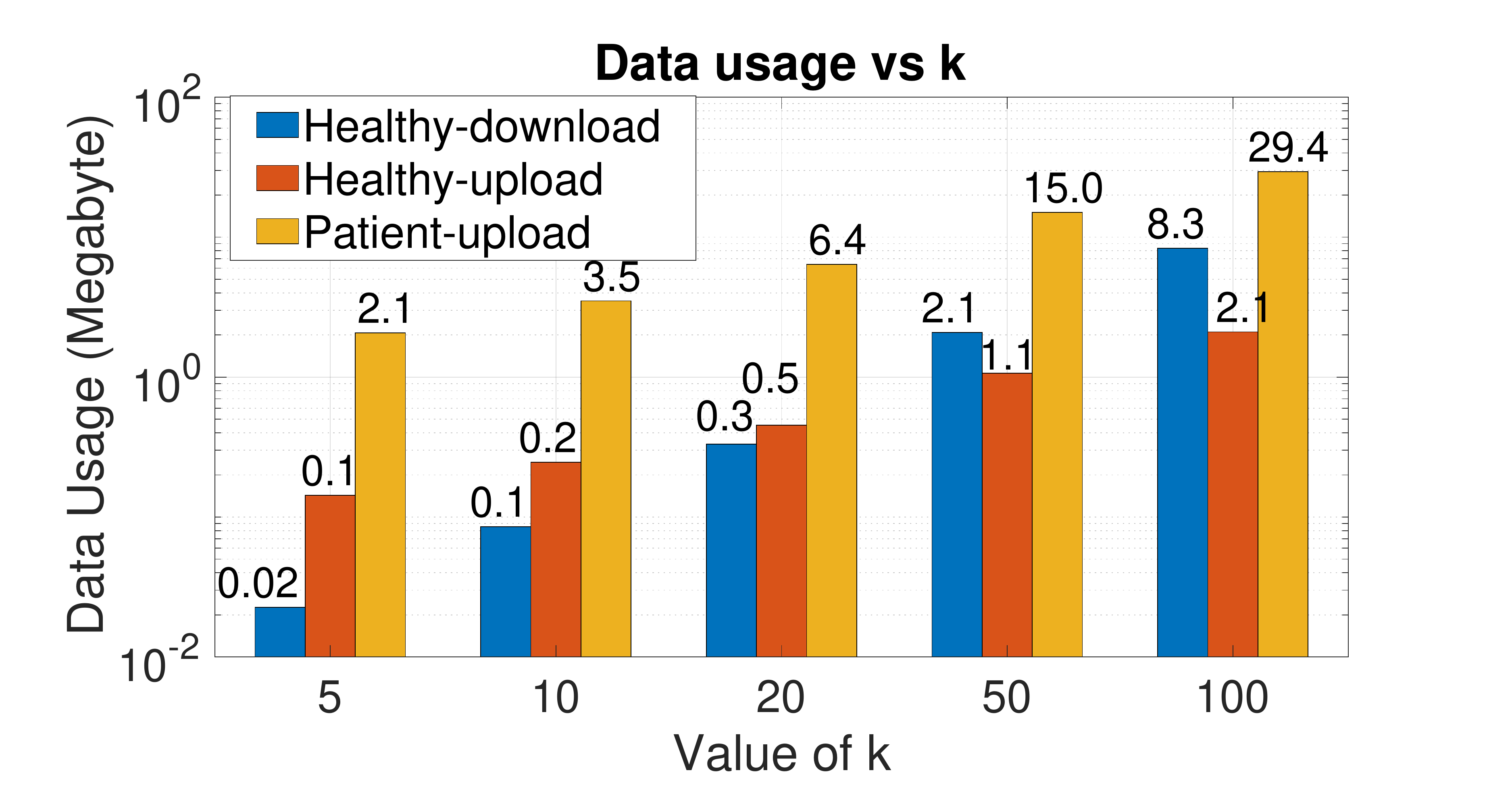}
    
    \caption{Communication overhead under different $k$ for (a) healthy user upload, (b) healthy user download, and (c) patient upload.  }
    \label{fig:commu-overhead}
\end{figure}

\paragraph{Communication Overhead} We evaluate the system's communication overhead from three perspectives: (1) the daily upstream data usage for a healthy user, (2) the daily downstream data usage for a healthy user, and (3) upstream data usage for a patient uploading records of past 14 days. We measured the data usages in each perspective with  $k$ ranging from 5 to 100. The measurement results are presented in Figure~\ref{fig:commu-overhead}.

\begin{sloppypar}
We have two key observations. First, the communication overhead is trivial for both healthy users and patients when $k$ is small. For example, when $k$ is 5, the total amount of exchanged data between a healthy user and \sysName servers is only 0.12 megabytes(MB) per day. Second, the communication overhead remains acceptable even with a large $k$. When $k=100$, the data usages for a healthy user and a patient are 10.4 MB and 29.4 MB, respectively.
\end{sloppypar}

The main takeaway from the above results is that users of \sysName app can adaptively change the value of $k$ based on the communication environment. For example, when the smartphone is connected to WiFi, a large $k$ can be used so that stronger privacy protection can be provided. When the smartphone's communication capacity is scarce, a smaller $k$ can be used to save the data usage while sacrificing some privacy protection. 

% For an infected user, the communication overhead comes from the messages to be uploaded. The uploaded messages are made up of the user’s encrypted trajectory and EphIDs, as shown in 1a and 1b in Figure~\ref{fig:ble-based}. For a normal user, we consider the communication cost from two perspectives: upload and download. The data a normal user needs to upload includes his past trajectories and geohash codes for querying \redzone (2a and 2b in Figure~\ref{fig:ble-based}), and masked encrypt exposure risks for querying \riskInter(4a and 4b in Figure~\ref{fig:ble-based}). For the data needs to be downloaded, we consider encrypted exposure risks and encrypted EphIDs from \redzone and the decrypted plain-text values from \riskInter. 

% As the communication overhead is proportional to the value of $k$ in k-anonymity, we measure the normal users' and patients' communication overheads by setting $k$ to different values. The measurement result is shown in Figure~\ref{fig:commu-overhead}. \todo{A paragraph describing the figure}. 

\paragraph{Computation Overhead} 

% In \sysName when both GPS and BLE are turned on, the smartphone's location and advertised EphID update per minute, then in an extreme situation, for the past 14 days, a user can have $14 days \times 24 hrs \times 60mins =  20160 $ location records and 20160 used EphIDs. Assume there are $k (k \in Z^+)$ sets of fake trajectories, then in total there will be $20160(k+1)$ location points and $20160(k+1)$ used EphIDs. The number of received EphIDs depends on how many people the user has met in the past, since fake EphIDs are assigned to each fake location, we estimate that the number of real and fake received EphIDs is also $20160(k+1)$. Based on the above calculation, communication overhead, as well as the following computation overhead of the \sysName system will be estimated. 
% \todo{add parameters to the Paillier alg}

\begin{figure}[t]
    \centering
    \includegraphics[width=1.0 \columnwidth]{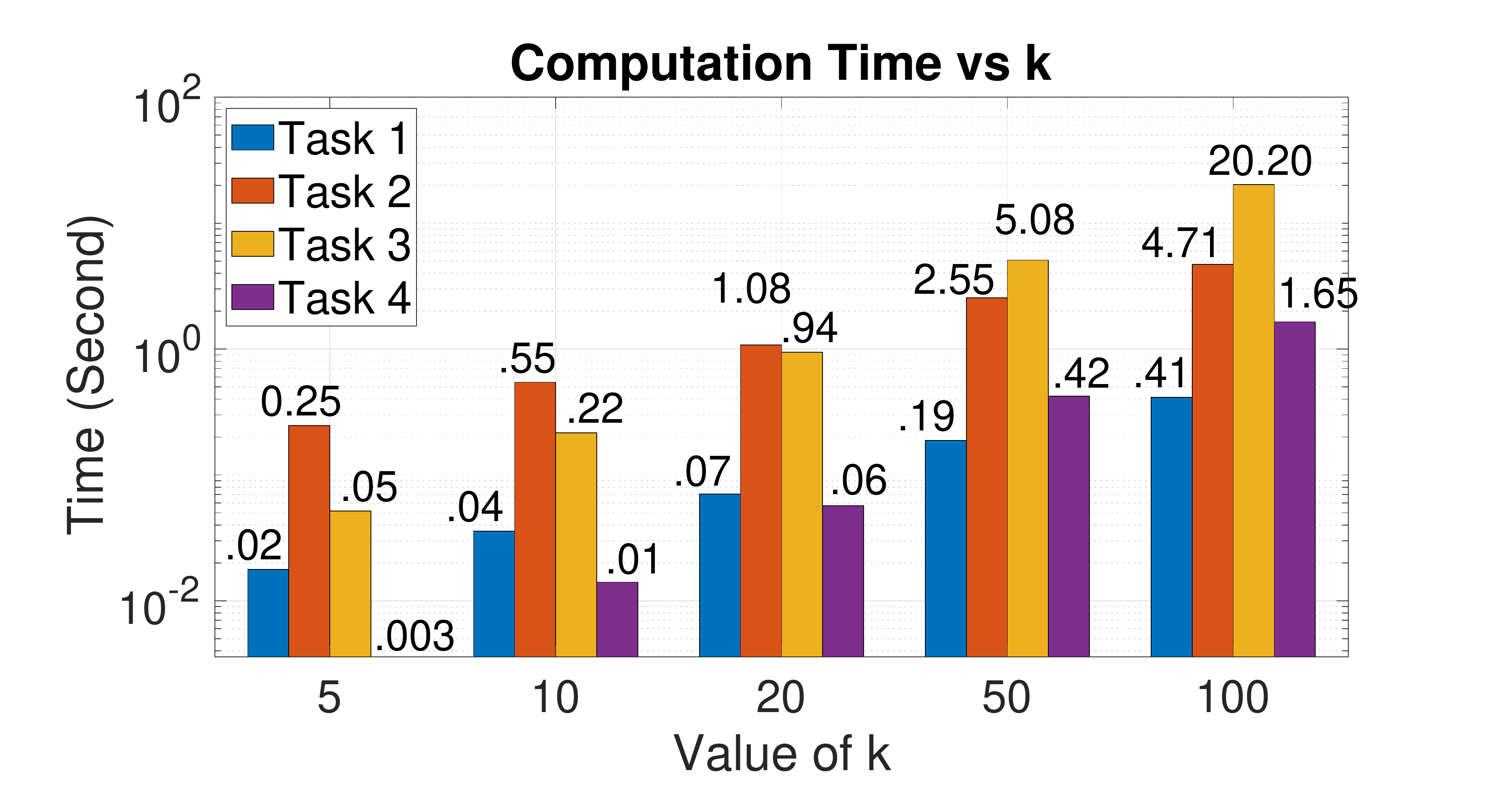}
    \caption{Computation time of (1) processing risk query result, (2) encrypting patient's data, (3) integrating patient data input and (4) computing exposure risk for a query.}
    \label{fig:compu-overhead}
    \vspace{-0.2in}
\end{figure}

\begin{table}[t]
\caption{Benchmark of Paillier Cryptosystem. Public key size: 2048 bits, private key size: 2048 bits.}
\centering
\resizebox{0.7\columnwidth}{!}{%
\begin{tabular}{|l|c|c|c|}
 \hline 
\                & \textbf{Encryption} & \textbf{Addition} & \textbf{Scaling} \\ \hline 
\textbf{Mobile} &               50.23 ms &             2.79 ms &                      4.39 ms 
% & 7.6ms  
\\ \hline 
\textbf{Desktop} &                  5.82 ms &              0.11 ms &                      0.12 ms 
% & 0.61 ms
\\ \hline 
 
\end{tabular}
}

\label{tab:pailliar-runtime}

\vspace{-0.2in}
\end{table}
The main computation overhead comes from the Paillier cryptosystem. The smartphone's operations involve Paillier encryption and addition, \redzone involves  Paillier encryption, addition, and scaling. Table~\ref{tab:pailliar-runtime} presents the benchmark of Paillier cryptosystem measured on a smartphone and a desktop separately. In addition, we measure the practical computation time for four tasks: (1) processing a risk query result at a healthy user's phone, (2) encrypting a two-week window of data at a patient's phone,  (3) integrating a patient's data input at \redzone, (4) exposure risk computation at \redzone.  The measurement results are shown in Figure~\ref{fig:compu-overhead}. 

From Figure~\ref{fig:compu-overhead}, we can observe that when a healthy user sends a query, the computation time of both task (1) and (4) is trivial which take less than 2 seconds even when $k=100$. In addition, task 2 and task 3 take longer time where patients' data are processed. It takes 4.71 seconds and 20.20 seconds for task 2 and task 3 when $k=100$, respectively. The reason that data integration takes longer time than other tasks is that it considers the dying down of the virus in an environment over time, which requires additional homomorphic scaling and addition computations when compared to other tasks.

% From Table \ref{tab:pailliar-runtime}, it takes RedZone Collector around 2.2 seconds to do homomorphic computation on each patient’s data. When a normal user query the database, RedZone needs to do homomorphic subtraction on each pair of matched patient’s EphID and normal user’s EphID, if there are $n$ matched pairs, the time for homomorphic subtraction will be $6.05n $ milliseconds. 

% \todo{The process to get the numbers in table 1 is not clear }
% \begin{table}[]
% \centering
% \resizebox{0.45\textwidth}{!}{%
% \begin{tabular}{@{}llcll@{}}
% \toprule
% \multicolumn{1}{c}{} & \multicolumn{1}{c}{\textbf{Computation}} &  \phantom{abc} & \multicolumn{2}{c}{\textbf{Communication}} \\ \cmidrule(l){2-2} \cmidrule(l){4-5} 
% \multicolumn{1}{c}{}   & \textbf{Runtime} && \textbf{Upload}      & \textbf{Download} \\ \midrule
% \textbf{Infected User} & 34(k+1) min      && 56(k+1)MB            & 0                 \\
% \textbf{Normal User}   & 2.79n ms         && 1.5(k+1) MB+ 1.4n KB & 2.8nKB            \\ \bottomrule
% \end{tabular}
% }
% \caption{Computation and communication overhead for users, $k$ represents $k$ sets of fake trajectories, $n$ is the number of intersections between query trajectories and patients' trajectories. \todo{update table}}
% \label{tab:my-table}
% \end{table}

\paragraph{Daily Battery Consumption}
% \sysName involves with both geo-location and Bluetooth tracing, and some encryption operations are also performed in the smartphone side. Therefore, \sysName will consume more battery power, comparing with other contact tracing apps.
We measure the daily battery consumption of the \sysName app. The daily battery consumption is dominated by geolocation logging and broadcasting Bluetooth EphIDs during the day. When setting the GNSS logging frequency to one log per 15 seconds,  the app consumed an averagely 439 mAh/day, which is roughly 12\% of the total battery.

It is important to note that the power consumption can be further reduced if we adapt the GNSS logging frequency based on different mobility patterns following methods similar to \cite{Kjrgaard2009}.  For example, based on Sila-Nowicka et al's work on human mobility~\cite{sila2016analysis}, a person spends $1.74$ hours in driving or public transporting, $0.66$ hours in walking, and $9.6$ hour in a stationary state. These mobility modes can be easily detected using IMU sensors in a smartphone\cite{Kjrgaard2009}.  Then,  we can adapt the GNSS logging frequency based on mobility modes, where 1 GNSS log is performed for every 15 seconds during walking, driving or public transporting period and 1 GNSS log is performed per 5 minutes during stationary mode. In our measurement, using this adaptive logging strategy, \sysName app consumes at average 121 mAh, which is 3.2\% of the total battery. 

We also measure the battery consumption at a smartphone for processing \sysName data. The measurement results are presented in Figure~\ref{fig:k-power} in Appendix. The results indicate that cryptography protocols consume trivial battery power.  When $k=100$, the power consumption for a healthy user is less than 0.5\% of total battery and less than 2\% for a patient.

\subsection{Comparison with other DCT schemes}
In this section, we compare the performance of \sysName with other existing DCT methods. Communication and computation overhead for most Bluetooth-based methods is negligible compared with \sysName. Take DP-3T as an example, the up-link size for a patient is less than 1KB, and down-link size for a healthy user varies from 72KB to 6MB assuming 2000 new cases a day~\cite{troncoso2020decentralized}. The computation overhead of DP-3T mainly comes from comparing EphIDs in plaintext, which is trivial compared with Paillier operations used in \sysName. For geolocation-based methods, most of current protocols did not provide implementations. Here we choose one geolocation-based method called covid-alert~\cite{covid-alert} for comparison, which uses PSI and Paillier cryptosystem with code provided. In our measurement, the up-link size for a patient in covid-alert is 2.5MB, the uplink size for an healthy user is 57.7MB, which is much larger than healthy users' uplink size in \sysName. Besides, it takes around 2.6 minutes for a healhty user's app to process all the data during a query. It is also worth pointing out that covid-alert does not protect patient privacy against untrusted servers.

% The app only logs in out-of-home period, hence we consider the logging time to be 12 hours a day. 

%  which the logging frequency would be $15$ seconds, $1$ minutes, $5$ minutes respectively. Apart of this normal routine, we also consider a busy routine that a user is constantly moving, like a delivery man, and the logging frequency would always be $15$ seconds. 

% Based on the two situations, we measure the power consumption from two aspect: the tracing consumption and privacy consumption, and will be measured by the android-build-in battery manager. 

% Tracing consumption includes the power consumed by geo-location and Bluetooth logging. 

% Privacy consumption includes the power consumption of fake trajectory generation and encryption operations. In this part, we separate the diagnosed users from the healthy users, who need to encrypt the trajectories of the pasted two weeks, and we evaluated the power consumption under different k-anonymity, which is shown in Figure~\ref{fig:k-power}. \todo{add fig and brief explain} Both the tracing power and the privacy power show \sysName have no problem in daily usage.

\section{Conclusions and Future Work}\label{sec:conclusion}
In this paper, we proposed \sysName to supplement manual contact tracing. It uses both geolocation-based and Bluetooth-based location data and protects all user privacy using k-anonymity, homomorphic encryption, and MPC. 
In the future, we plan to enhance its accuracy and interoperability in several ways. 
First, other types of location data may be incorporated with ease because our privacy protection scheme and exposure risk computation techniques are independent of the type of location data used.
We provide an example of integrating WiFi access point based location data to \sysName in Appendix~\ref{sec:Appendix-1}. 
Second, we plan to use open-source transportation mode detection(TMD) tools ~\cite{tmdGoogle} and IMU and/or GNSS speed sensors in mobile phones to enhance the accuracy of \sysName app. 
Since an app may be in motion when the app logs the contact information, detecting the mode of transportation will aid in accurate location logging and exposure risk computation. 
Third, we plan to design a dedicated interface for adding authorized data from manual contact tracing and utilizing the patients' location data from the \redzone. 
The former will help \sysName app users who did not personally know the patients without \sysName app by allowing manual contact tracers to add such patients' past geolocation data to the \redzone. 
The latter will provide recommendations to health authorities for zonal lockdown instead of a statewide lockdown by identifying potential hot-spots and super-spreader events from the patients' location data in the \redzone. 
With these added features, we plan to perform a small-scale study of \sysName app on a university campus. 
We also plan to open-source the code for \sysName after the review process. 

\begin{comment}
\subsection{Regional Cooperation}
 \sysName~ maybe effectively operated across regional borders to enable freedom of travel in this global era. At setup, a \redzone is registered with a particular geographic region, such as \textbf{country/state/city}. The physical boundary of each \redzone's operation is determined by the geolocation coordinates (latitude and longitude) and its corresponding geohash. 
 When \sysName~ is active, if a \redzone~ receives uninfected user or patient trajectories and EphIDs from areas outside its boundary, it will forward them to the \redzone operating in that area. 
 The concrete details of trajectory forwarding to the correct sever must be decided among participating regions. 
 In a similar design,  ROBERT~\cite{robert20} assigns a country code and adds the encrypted country code as a part of the broadcasted message. 
 The regional integration is an advantage to \sysName~ which is not easily achievable in Bluetooth-based systems, particularly decentralized systems.

\end{comment}

% \begin{acks}
% \end{acks}

% \newpage

\bibliographystyle{ACM-Reference-Format}
\bibliography{main}

\newpage

\appendix

\section{Appendix: Other supporting materials}

\begin{figure}[h]
    \centering
    \includegraphics[width=0.85 \columnwidth]{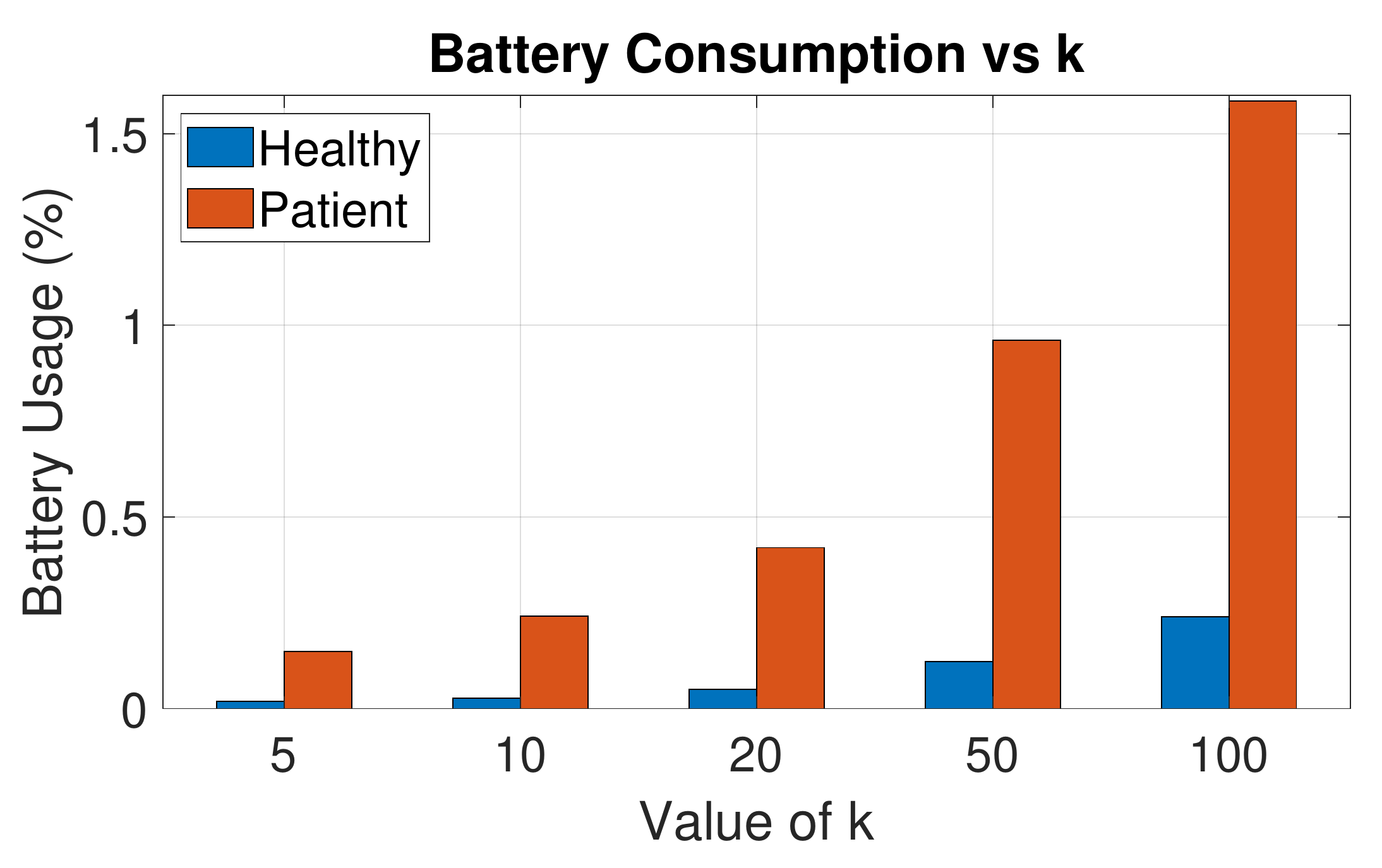}
    \caption{Battery consumption for encryption operation under various $k$ values}
    \label{fig:k-power}
\end{figure}

We put other supplementary materials in this section. Figure~\ref{fig:k-power} shows the measurement results of a smartphone's battery consumption for processing the data from a patient and a healthy user, respectively. Note that the app can choose to process the data only if it is charged. Algorithm~\ref{alg:paillier} describes the Paillier cryptosystem which is used in \sysName. Table~\ref{tab:my_symbol} lists the symbols used for describing \sysName.

% encryption operations on the smartphone side. For healthy users, the retrieved cipher-text need Paillier subtraction before sending to the Risk Interpreter, which is typically happened in night. For patients, who is diagnosed and quarantine at home, Paillier algorithm is needed to encryption the trajectories in the pasted 14 days before uploading to \redzone. 

\begin{algorithm}[h]
 \caption{Paillier Crypto System}
\SetAlgoLined
\keyGeneration{}
 1. Choose two large random prime number $p$ and $q$, ensuring that $gcd(pq,(p-1)(q-1))=1$\;
 2. Compute $n=pq, \lambda = lcm(p-1,q-1)$ \;
 3. Choose random integer $g$ where $g \in \mathbb{Z}^*_{n^2}$, compute $\mu = (L(g^\lambda \; \text{mod} \; n^2))^{-1} \; \text{mod} \; n$, where $L(x) = \frac{x-1}{n}$ \;
 4. The public key is (n,g), and the secret key is $(\lambda,\mu)$\;
 
 \encryption{}
 1. Let $m$ is a message to be encrypted where $m\in \mathbb{Z}_n$ \;
 2. Select random $\gamma$ in $m\in \mathbb{Z}_n$\;
 3. Encrypt $m$ and get its ciphertext $\llbracket m \rrbracket$ by $\llbracket m \rrbracket=E(m,\gamma) = g^{(m+n\gamma)} \; \text{mod} \; n^2$\; 
 \decryption{}
 1. Decrypt $\llbracket m \rrbracket$ and get its plaintext $m$ by $m = L(\llbracket m \rrbracket^\lambda \; \text{mod} \; n^2) \cdot \mu \; \text{mod} \; n$ \; 
 \homoPro{}
\textbf{\textit{Addition:}} (notation: $\llbracket m_1 \rrbracket \oplus \llbracket m_2 \rrbracket$) \newline
$E(m_1,\gamma_1)E(m_2,\gamma_2) \text{ mod } n^2 = E(m_1+m_2, \gamma_1+ \gamma_2) \text{ mod } n^2$ 

\textbf{Scale:} (notation: $m_2 \otimes \llbracket m_1 \rrbracket $) \newline
$E(m_1,\gamma_1)^{m_2} \text{ mod } n^2 = E(m_1 \cdot m_2, \gamma_1 \cdot m_2) \text{ mod } n^2$ 

% \textbf{Subtraction:} (notation:  $\llbracket m_1 \rrbracket \ominus \llbracket m_2 \rrbracket$)\newline
% $[E(m_1, \gamma_1)[E(m_2,\gamma_2)^{-1} \text{ mod } n^2]] \text{ mod } n^2 = E(m_1-m_2, \gamma_1 - \gamma_2) \text{ mod } n^2$

 \label{alg:paillier}
\end{algorithm}

\begin{table}[t]
\caption{Symbol and definition.}
    \resizebox{1.0\columnwidth}{!}{
    \centering
    % \ra{1.3}
    \footnotesize
    \begin{tabular}{|c|p{6.2cm}|}
        
        \hline
        \textbf{Symbol} & \textbf{Definition} \\
        \hline
        
        $(l_i,t_i)$ & A location tuple indicate location $l_i$ is visited at time $t_i$. \\
        \hline 
        
        $\bm{L_r}$ & A GNSS trajectory recorded by the user's smartphone, $\bm{L_r} = \{(l_i,t_i) | i \in Z^+\}$. \\
        \hline 
        
        $\bm{ L_{fi}}$ & A fake trajectory intimating the real trajectory's mobility pattern. \\
        \hline
        $\bm{\tilde{L}}$ & A trajectory superset containing both real and fake trajectories, $\tilde{\bm{L}} = \{\bm{L_r}, \bm{L_{f1}},\bm{L_{f2}}...\bm{L_{fk}} \} $. \\ 
        \hline 
        $s_i$ & A flag indicating the reality of $(l_i,t_i)$, $s_i = \{0,1\}$. \\ 
        \hline 
        
        $\llbracket . \rrbracket$ & The encryption function in Paillier cryptosystem.\\ 
        \hline 
        
        $r_{lt}$ & A flag in \redzone indicating how many patients visited $l$ at time $t$. \\
        \hline
        $\bm{R}$ & $ \bm{R}  := \{r_{lt} | \textit{ for all possible } l \textit { and } t<2 \; weeks \}$ \\
        
        \hline 
        
        $W_{lt}$ & Exposure risk for location $l$ at time $t$ \\
        \hline
        
        $W$ & Exposure risk for a user, calculated based his/her past trajectories \\
        \hline
        
        $g_i$ &  Geo hase code at time $t_i$ \\
        \hline
        
        $u_i$ & Unique random string embedded in Bluetooth beacon at time $t_i$ \\
        \hline
        
        $\tilde{\bm{E}}$ & Bluetooth contacts sent by the patients, cloaked with fake ones, $\tilde{\bm{E}} = \{(g_i, t_i, u_i)|i \in Z^+ \}$ \\
        \hline
        
        $\bm{E_r^{Q}}$ & Real Bluetooth contacts sent to the server for query, $\bm{E_r^{Q}}$ = $\{(g_i, t_i)|i \in Z^+ \}$ \\
        \hline
        
        $\bm{E_F^{Q}}$ & Fake Bluetooth contacts for query \\
        \hline
        
        $\bm{E^{Q}}$ &  A Bluetooth contact superset for query, $\bm{E^{Q}}$ = $\{\bm{E_r^{Q}}, \bm{E_F^{Q}} \}$ \\
        \hline
        
        $m$ & Bluetooth beacon comparison result by subtracting $u_i$ from patients and healthy users \\
        \hline
        
        \end{tabular}
    \vspace{-0.8in}
    }
    
    \label{tab:my_symbol}
    \vspace{-0.2in}
\end{table}

\section{Appendix: Enhancing Accuracy with WiFi Access Point based location data}\label{sec:Appendix-1}

As mentioned in Section~\ref{sec:conclusion}, \sysName is not restricted using to just Bluetooth or geolocation based location. 
The privacy protection scheme and exposure notification computations are generic and may be applied to any location data including WiFi access points IDs and ultrasound measurements~\cite{Novid}. 
In this appendix, we provide an example of integrating MAC addresses from WiFi access points as the third location data type to \sysName. 
A similar approach may be used for interoperability with other location data. 
% \sysName app obtains its user location from multiple sources, including GNSS sensors, neighboring WiFi access points’ SSID+MAC addresses, and neighboring cellular base station id\cite{}.
% Smartphone programming APIs for access such information are widely available \cite{}. 
Since the GNSS signals are usually weak indoor, neighboring WiFi access points’ MAC addresses are also recorded in addition to users’ longitude and latitude using existing smartphone programming APIs. 
% This increases the accuracy of \sysName’s trace intersection.

Using the example described in Section~\ref{section:geo-method}, Alice’s location trajectory becomes 
$\bm{L}:=\{(l_i,M_i, t_i) | i \in \bm{Z}^+ \}$, 
where $M_i$ is a set of received MAC addresses at time $t_i$: $M_i = \{m_1, m_2 \dots m_n\}$.
During the data encryption phase, fake MAC addresses are generated from a public geo-mac-address databases~\cite{WiGLE} to match the fake trajectories, $\bm{L_{fi}}$. 
All the real and fake MAC addresses are hashed using a one-way hash function, denoted by $hash(m)$. The encrypted trajectory data becomes 
$\llbracket \bm{S} \rrbracket:=\{(\llbracket s_{lt} \rrbracket ,l,t, hash(M_{lt}))| \forall (l, t)\in \tilde{\bm{ L}}\}$, 
where $ hash(M_{lt}) = \{hash(m_{lt1}),hash(m_{lt2}) \dots hash(m_{ltn}) \}$.
Note that only WiFi mac addresses with strong signal strength will be recorded to reduce false positives in exposure risk computation.

The \redzone creates a new MAC-address risk flag through homomorphic addition: $ \llbracket r\rrbracket_{mt} =\llbracket r \rrbracket_{mt} \oplus \llbracket s \rrbracket_{lt} $, similar to  $\llbracket r \rrbracket_{lt}$, $\llbracket r \rrbracket_{mt}= \llbracket k\rrbracket  \textit{ for some } k\geq1 $.
The flag indicates if confirmed COVID-19 patient(s) have been in the range of WiFi access point $m$ at time $t$. 
As in Section~\ref{subsec:geo-exp-risk-Q}, Bob queries the \redzone for his exposure risk with his cloaked superset of real and fake trajectories with their corresponding MAC addresses. 
In addition to replying with the exposure risks for geolocation ($ \llbracket \bm{W} \rrbracket$), the \redzone also replies to Bob with a  list of exposure risks at WiFi access point $m$ at time $t$, $ \llbracket U \rrbracket_{mt} := \{ \llbracket r \rrbracket_{m\tau} | t \leq  \tau \leq t-\delta \}$.

The privacy of \sysName is maintained using the original homomorphic encryption, k-anonymity, and MPC based protection scheme. 
By using a public database to generate MAC addresses for fake trajectories, we ensure that k-anonymity is maintained with the addition of a new type of location data. 
Thus, the addition of neighboring WiFi access point's MAC address enhances the accuracy of geolocation-based DCT in indoor environments without compromising \sysName's privacy. 

% By enhancing with the neighboring WiFi access points, the location-based contact tracing can be realized both indoor and outdoor. 
% What’s more, the way that fake mac addresses are generated and hashed also makes sure the RedZone collector can’t decrypt users' trajectories, even if it has a precise geo-mac-address database for comparison.  

\section{Appendix: Supplementary Security Analysis}
\label{sec:supp_secu_analysis}
\revisionok{We describe the rest of the attacks in Table~\ref{tab:security_table}}
\subsection{Relay and Replay Attack }

% \paragraph{Replay Attack:}
One problem with Bluetooth-based methods is that an attacker may collect advertised EphIDs from a patient in one place (e.g. hospital) and broadcast them in other places, causing devices received the replayed EphIds to later falsely believe they met patients. Relay attacks occur when the EphIDs are re-broadcasted immediately at a different location, while replay attacks occur when EphIDs are collected and broadcasted at different times. Some of the existing works~\cite{blueTraceWhitepaper,troncoso2020decentralized,googleAppleDCT} integrate time information into EphIDs and validate if the received EphIDs are in the correct time slots to prevent replay attacks. 
Relay attacks may be prevented by using geolocation data along with EphIDs~\cite{reichert2020survey,gvili2020security}. 
% Nevertheless, our system is protected from both relay and replay attack.
Our EphIDs contain both the timestamp and geo hash which detects both relay and replay attacks. 
By checking the geohash code from a received EphID, the \sysName app can obtain the sender's rough location and discard EphIDs whose locations are too far away from the receiver.

\subsection{Aggressive Broadcasting Attack}

% \paragraph{Bluetooth \& Navigation Attacks:} 
 As most of the current DCT apps rely on Bluetooth technology, an attacker may attempt to stop DCT apps functioning in specific premises by disrupting Bluetooth broadcasting: 
\begin{itemize}
    \item \emph{Bluetooth jamming:} The attacker can undermine the BLE tracing function with a Bluetooth jammer to block the exchange of EphIDs in a certain area.
    \item \emph{Bluetooth overwhelming and storage draining:} The attacker can deploy plenty of low-cost BLE devices and broadcast massive EphIDs. This can drain up the storage and energy of the phone. 
    % \item \emph{Bluetooth injection} Bluetooth protocol is not perfectly secure that many venerability exists. 
    % \item \emph{Navigation spoofing} The navigation system in a smartphone is most depends on Global Navigation Satellite System (GNSS). However, a GNSS spoofer~\cite{zeng2018all} can imitate the satellites' signals to deliver the fraud geo-location information to the device. Moreover, if the phone is hijacked, it is easy to change the system geo-location~\cite{lexaFakeGPS}. This attack can raise both the false positive and false negative detection of our system.
\end{itemize}

The above attacks may force app users to turn off Bluetooth in their smartphone. Bluetooth-only DCT apps thus cannot recover from these attacks. However, turning off Bluetooth will not stop \sysName from working as it also uses localization techniques for contact tracing. One may argue that GNSS signals can also be jammed and messed with. However, smartphones do not only rely on GNSS signals for localization but also use WiFi and cellular networks. The cost for disrupting smartphone localization is much higher than achieving the same in Bluetooth. 

\subsection{Data Poisoning Attack}
% \paragraph{Data Poisoning Attack:}

Data poisoning attack~\cite{steinhardt2017certified} occurs when an attacker adds malicious data like false trajectories to \redzone. \sysName can prevent data poisoning attack by considering three types of attackers. First, for the attacker who is uninfected but claims to be a patient, certificates have been utilized to ensure only patients verified by health authorities are allowed to upload their data to the \redzone. The patients who did not install the app before diagnosis and who are willing to manually upload their visited places will be assisted by their contact tracers, as mentioned in section~\ref{sec:geo_data_collect}. We assume the communications between patients and contact tracers are honest, which is a fundamental assumption in manual contact tracing. Second, for the attacker who is an infected user with the secret code but attempts to upload incorrect data to \redzone, the data modification is avoided as we use secure storage to store data on the phone~\cite{androidSecurityData, iosSecurityFile}. Third, for the attacker who is uninfected and query with incorrect data, either trajectories or EphIDs from health users will not be used for computing others' exposure risks, thus these incorrect data cannot pollute \redzone. 

\subsection{Man in the Middle Attack}
% \paragraph{Man in the Middle Attack:} 

This attack occurs when the app and servers' communications are not authenticated. In this scheme, an attacker can alter a patient's trajectory data sent to the RedZone Collector or modify the plaintext risk level returned from the Risk Interpreter. However, in the proposed system we ensure all the communications between the app and servers are mutually authenticated and encrypted through TLS channel, which prevents this attack. 

% \input{discussion}

% \input{survey}

%%%%%%%%%%%%%%%%%%%%%%%%%%%%%%%%%%%%%%%%%%%%%%%%%%%%%%%%%%%%%%%%%%%%%%%%%%%%%%%%
\end{document}